\pdfoutput=1
\documentclass[aps,prd,floatfix,amsmath,amssymb,onecolumn,groupedaddress,11pt]{revtex4}
\usepackage{graphicx}
\usepackage{amssymb}
\usepackage{epstopdf}
\usepackage{subfigure}
\usepackage{array}
\usepackage{color}
\usepackage{pdfsync}
\usepackage{amsmath}
\usepackage{hyperref}
\newcommand{\st}[0]{\mathcal{S}_{3}}

\begin{document} 

\title{Sizable Stop Mixing in Flavored Gauge Mediation Models with Discrete Non-Abelian Symmetries} 
\author{Lisa L.~Everett}
\email{leverett@wisc.edu} 
\author{Todd S.~Garon}
\email{tgaron@wisc.edu}  
\author{Ariel B.~Rock}
\email{arock3@wisc.edu}
\affiliation{Department of Physics, University of Wisconsin-Madison, Madison, WI 53706}
\date{\today}

\begin{abstract}
We analyze a minimal flavored gauge mediation model in which the electroweak Higgs and messenger doublets are embedded in multiplets of a discrete non-Abelian symmetry.  In this scenario, the minimal Higgs-messenger sector that is consistent with the 125 GeV Higgs mass has two vectorlike pairs of messenger fields. This scenario is obtained in a specific limit of the superpotential interactions of the Higgs-messenger fields and the matter fields. Due to the structure of the messenger-matter Yukawa couplings in this limit, sizable stop mixing and flavor-diagonal soft supersymmetry breaking parameters are achieved. In most of the parameter space, the masses of the colored superpartners are at most in the $5-6$ TeV range.

\end{abstract}
\maketitle 

\section{Introduction}
The 2012 discovery of the 125 GeV Higgs particle \cite{Aad:2012tfa,Chatrchyan:2012xdj} and subsequent detailed measurements of its properties at the Large Hadron Collider (LHC) has provided significant limits on the allowed possibilities for extensions of the Standard Model (SM).  In the context of theories with softly broken supersymmetry at the TeV scale (for reviews, see e.g.~\cite{Martin:1997ns,Chung:2003fi}), the Higgs mass is known to be within the theoretically allowed range for perturbative theories, but its relatively high value either requires large radiative corrections in the minimal supersymmetric standard model (MSSM), or an enlarged Higgs sector to boost the tree-level contributions. As such, it has long been known in the MSSM that large stop mixing or very heavy stops are needed (see e.g.~\cite{Carena:2002es}). This can place stringent constraints on specific models of the soft supersymmetry breaking terms, and also has important implications for the potential observability of superpartners at the LHC.

The model-building constraints imposed by the Higgs measurements are particularly striking in the context of gauge mediation. In its minimal implementation, gauge-mediated supersymmetry breaking \cite{gauge1,gauge2,gauge3,Giudice:1998bp} predicts highly suppressed scalar trilinear couplings ($A$ terms) at the messenger scale, leading to small stop mixing.  As a result, consistency with the Higgs data \cite{Draper:2011aa,Arbey:2011ab,Ajaib:2012vc} generally requires very heavy $SU(3)$-charged superpartner masses and/or high values of the messenger scale.  This conclusion can be circumvented if the messenger fields can couple directly to the MSSM fields, as explored in  \cite{gauge3,Giudice:1998bp,Chacko:2001km,Shadmi:2011hs,Evans:2011bea, Evans:2011uq,Evans:2012hg,Kang:2012ra,Craig:2012xp,Albaid:2012qk,Abdullah:2012tq,Perez:2012mj,Byakti:2013ti,Evans:2013kxa,Calibbi:2013mka,Evans:2015swa,Galon:2013jba,Fischler:2013tva,Calibbi:2014yha,Ierushalmi:2016axs}
Of particular interest are the ``flavored gauge mediation'' models   \cite{Shadmi:2011hs,Abdullah:2012tq,Perez:2012mj,Byakti:2013ti,Evans:2013kxa,Calibbi:2013mka,Evans:2015swa,Galon:2013jba,Fischler:2013tva,Calibbi:2014yha,Jelinski:2015voa,Ierushalmi:2016axs,Everett:2016meb,Ahmed:2016lkh}),  
for which there is nontrivial mixing of the $SU(2)$ messenger doublets and the MSSM Higgs fields.  These models allow for the generation of one-loop $A$ terms at the messenger scale, alleviating the Higgs mass problem of minimal gauge mediation in the MSSM.

In building models of flavored gauge mediation, an underlying Higgs-messenger symmetry is typically employed to control the mixing of these fields. A logical and now-standard choice is to use a $U(1)$ symmetry as the Higgs-messenger symmetry, and many interesting examples of this type have been proposed in the literature (see e.g.~\cite{Ierushalmi:2016axs} for a recent analysis and set of LHC benchmark points). In this case, the $U(1)$ charges are chosen judiciously to control the couplings of the Higgs and messenger fields so as to obtain nontrivial third generation $A$ terms and to avoid generating dangerous interactions between the MSSM Higgs fields and the supersymmetry breaking sector.

An alternative is to choose a discrete non-Abelian symmetry as the Higgs-messenger symmetry, as proposed in \cite{Perez:2012mj}.   
This idea was studied using the specific choice of $\mathcal{S}_3$, the permutation group on three objects, for the case of two families in \cite{Perez:2012mj}, and extended to three families in \cite{Everett:2016meb}. In these analyses, it was shown that if the SM quarks and leptons transform nontrivially with respect to the $\mathcal{S}_3$ symmetry (as  required for at least a subset of the SM matter particles), obtaining nontrivial SM Yukawa coupling entries in the diagonal fermion mass basis led to vanishing entries in the corresponding messenger Yukawa couplings, as a direct consequence of the non-Abelian symmetry.  As a result, generating the needed large top quark Yukawa couplings thus led to vanishing top quark messenger Yukawa couplings, unless the relevant fields are all taken to be $\mathcal{S}_3$ singlets.  The phenomenological implications of this correlation are that the resulting stop mixing is generically very small, such the superpartner masses must be quite heavy to be consistent with the Higgs data.

In this work, we analyze a three-family flavored gauge mediation model based on the $\mathcal{S}_3$ Higgs-messenger symmetry in which sizable third generation Yukawa couplings to both the Higgs and the messengers can be simultaneously generated.  We show that a fermion mass hierarchy and flavor-diagonal messenger Yukawa coupling structure can emerge in a specific limit of the renormalizable superpotential interactions of these fields that can result from additional symmetries placed on the system.  The resulting model of the soft supersymmetry breaking parameters is a minimal scenario with two pairs of vectorlike messenger fields.  As this yields larger stop mixing than was possible in previously studied examples of this type with renormalizable couplings of the MSSM fields and the Higgs-messenger fieds only  \cite{Perez:2012mj,Everett:2016meb}, the superpartner masses can be significantly lighter, with the heaviest superpartners at or below 5-6 TeV.

The structure of the paper is as follows.  In the next section, we present a brief overview of the general model framework as well as the detailed model inputs of this specific scenario based on the choice of $\mathcal{S}_3$ as the Higgs-messenger symmetry.  We then present the resulting soft supersymmetry parameters and carry out a numerical analysis of the superpartner spectra. Finally, we summarize and discuss prospects for future model-building directions along these lines.

\section{Model}
\label{backgroundsection}
In this section, we provide a self-contained review of the model framework and present the details of the specific scenario that is the focus of this paper. 

As a prelude, we note several salient features of the group theory of $\mathcal{S}_3$, as can be found in many references (see e.g.~\cite{Perez:2012mj}).  The irreducible representations of $\mathcal{S}_3$ are the singlet $\mathbf{1}$, a one-dimensional representation $\mathbf{1}^\prime$, and a doublet, $\mathbf{2}$, with tensor products 
\begin{eqnarray}
\mathbf{1}\otimes \mathbf{2}=\mathbf{2}, \qquad \mathbf{1}^\prime \otimes  \mathbf{2}=\mathbf{2}, \qquad  \mathbf{2}\otimes \mathbf{2}=\mathbf{1}\oplus \mathbf{1}^\prime\oplus\mathbf{2}.
\end{eqnarray}
As in \cite{Perez:2012mj} and \cite{Everett:2016meb}, all fields will be taken to be either singlet or doublet representations.  We use the basis in which the singlets obtained from the tensor products of two or three doublets are
\begin{eqnarray}
(\mathbf{2} \otimes \mathbf{2})_\mathbf{1}&=& \left [\left (\begin{array}{c} a_1 \\ a_2 \end{array} \right ) \otimes  \left (\begin{array}{c} b_1 \\ b_2 \end{array} \right ) \right ]_\mathbf{1} = a_1 b_2 +a_2b_1. \\
(\mathbf{2} \otimes \mathbf{2}\otimes \mathbf{2})_\mathbf{1}&=& \left [\left (\begin{array}{c} a_1 \\ a_2 \end{array} \right ) \otimes  \left (\begin{array}{c} b_1 \\ b_2 \end{array} \right )\otimes  \left (\begin{array}{c} c_1 \\ c_2 \end{array} \right ) \right ]_\mathbf{1} =a_1b_1c_1+a_2b_2 c_2.
\label{cg}
\end{eqnarray}

In this model framework, the minimal viable Higgs-messenger sector consists of one $\mathcal{S}_3$ doublet  $\mathcal{H}^{(2)}_{u,d}$ and one $\mathcal{S}_3$ singlet $\mathcal{H}^{(1)}_{u,d}$ for the up-type and down-type Higgs-messenger fields.  This is to stave off an otherwise severe $\mu/B_\mu$ problem (see \cite{Dvali:1996cu,Giudice:2007ca,Polonsky:1999qd} for a discussion of this issue within gauge mediation models), as discussed shortly.  Taking $\mathcal{H}_{u}=(\mathcal{H}^{(2)}_{u}, \mathcal{H}^{(1)}_{u})=(\mathcal{H}_{u1},\mathcal{H}_{u2},\mathcal{H}_{u3}) $ (and analogously for $u\rightarrow d$), we have 
\begin{eqnarray}
\mathcal{H}_{u}= \left (\begin{array}{c} \mathcal{H}_{u1}\\ \mathcal{H}_{u2}\\ \mathcal{H}_{u3}\end{array} \right ) = \mathcal{R}_u \left (\begin{array}{c} H_u\\M_{u1} \\  M_{u2} \end{array} \right ), \qquad  \mathcal{H}_{d}= \left (\begin{array}{c} \mathcal{H}_{d1}\\ \mathcal{H}_{d2}\\  \mathcal{H}_{d3} \end{array} \right ) = \mathcal{R}_d \left (\begin{array}{c} H_d\\M_{d1}  \\ M_{d2} \end{array} \right ),
\label{higgs_s3}
\end{eqnarray}
in which the electroweak Higgs fields are denoted as usual by $H_{u,d}$, the $SU(2)$ doublet messengers are  $M_{ui,di}$ ($i=1,2$), and $\mathcal{R}_{u/d}$ are unitary matrices that result from the diagonalization of the Higgs-messenger mass matrices (more on this shortly). We also have $SU(3)$ triplet messengers $T_{ui,di}$ ($i=1,2$) that are $\mathcal{S}_3$ singlets. The $T_{ui,di}$ and the messenger doublets $M_{ui,di}$ together form two pairs of $\mathbf{5}$, $\overline{\mathbf{5}}$ representations of $SU(5)$ (i.e., the number of messenger pairs $N_5=2$).

The model also includes two supersymmetry breaking fields: the $\mathcal{S}_3$ doublet $X_H$ and a $\mathcal{S}_3$ singlet chiral superfield $X_T$, where $X_H$ couples only to the Higgs-messenger doublets and $X_T$ couples only to   the triplet messengers. We assume that $X_T$ couples only to the triplet messengers and $X_H$ couples only to the messenger doublets or the MSSM fields, as needed to avoid the possibility of rapid proton decay (this requires additional symmetries, which is not difficult to implement in a concrete scenario) \footnote{We also note for that the different treatment of the doublet and triplet sectors undoubtedly poses challenges for any serious attempt to embed this scenario within grand unification; a thorough treatment of this question is beyond the scope of the present work.}. This field content and the relevant $\mathcal{S}_3$ charges are given in Table~\ref{tab:11}. 
\begin{table}[htbp]
   \centering
    \begin{tabular}{c|cccccc|cc}
     & $\mathcal{H}_u^{(2)}$&$\mathcal{H}_u^{(1)}$ & $\mathcal{H}_d^{(2)}$& $\mathcal{H}_d^{(1)}$ & $T_{ui}$ & $T_{di}$   &$X_H$ & $X_T$\\
    \hline
    $\mathcal{S}_3$ &$\mathbf 2 $& $\mathbf 1$& $\mathbf 2 $& $\mathbf 1 $ & $\mathbf 1 $ & $\mathbf 1 $ &\textbf 2 & $\mathbf 1 $\\
   \end{tabular}
   \caption{The $\mathcal{S}_3$ charges for the extended Higgs-messenger model described in this section.}
   \label{tab:11}
\end{table} 

The superpotential couplings of $X_H$ to the Higgs-messenger sector are given by
\begin{eqnarray}
W_H= \lambda X_H\mathcal{H}^{(2)}_u \mathcal{H}^{(2)}_d+\lambda' X_H \mathcal{H}^{(1)}_u \mathcal{H}^{(2)}_d+\lambda'' X_H \mathcal{H}^{(2)}_u \mathcal{H}^{(1)}_d +  \kappa M \mathcal{H}^{(2)}_u \mathcal{H}^{(2)}_d+  \kappa^\prime M \mathcal{H}^{(1)}_u \mathcal{H}^{(1)}_d.
\label{wh}
\end{eqnarray}
Here we will assume all couplings are real, and take $\lambda'=\lambda''=\lambda$ for simplicity. The supersymmetry-breaking field $X_H$ is then parametrized as follows:
\begin{eqnarray}
\langle  X_H \rangle = M  \left (\begin{array}{c} \sin\phi \\ \cos\phi \end{array} \right ) +\theta^2 F  \left (\begin{array}{c} \sin\xi \\ \cos\xi \end{array} \right ),
\label{deltavev}
\end{eqnarray}
where $\phi$ and $\xi$ characterize the vev directions of the scalar and $F$ components, respectively, and we take $F\ll M^2$.  After symmetry breaking, the effective superpotential takes the following form:
\begin{eqnarray}
W_H&=&  M \mathcal H_u^T \left(\begin{matrix}\sin\phi&\kappa &\cos\phi\\ \kappa &\cos\phi &\sin\phi\\ \cos\phi & \sin\phi &\kappa^\prime \end{matrix}\right)\mathcal H_d  + \theta^2 F \mathcal H_u^T\left(\begin{matrix}\sin\xi&0&\cos\xi\\ 0 &\cos\xi &\sin\xi\\  \cos\xi &\sin\xi & 0 \end{matrix}\right)\mathcal H_d \nonumber \\
&\equiv& \mathcal H_u^T\mathbb{M} \mathcal H_d+\theta^2 \mathcal{H}_u^T \mathbb{F} \mathcal{H}_d.
\label{extendedS3}
\end{eqnarray}
As outlined in \cite{Perez:2012mj}, we require $[\mathbb{M},\mathbb{F}]=0$, such that $\mathbb{M}$ and $\mathbb{F}$ are diagonalized by the same unitary transformation.  It is also necessary to have hierarchy of eigenvalues for both $\mathbb{M}$ and $\mathbb{F}$, to distinguish the MSSM Higgs fields from the messenger fields. Simultaneously incorporating both constraints is the underlying reason why we need to include the $\mathcal{S}_3$ singlet $\mathcal{H}_{u,d}^{(1)}$ in the Higgs-messenger sector within this framework, as these two conditions are incompatible if only $\mathcal{H}_{u,d}^{(2)}$ is included.

As shown in \cite{Everett:2016meb} for the case of Eq.~(\ref{extendedS3}), upon imposing $[\mathbb{M},\mathbb{F}]=0$, which requires $\kappa^\prime = \kappa = \sin(\phi-\xi)/(\cos \xi-\sin\xi)$ (for $\xi\neq \pi/4$),   
a viable solution with a distinct hierarchy of eigenvalues for both $\mathbb{M}$ and $\mathbb{F}$ can be obtained.  This distinct hierarchy is needed for the possibility of separate fine-tunings of the $\mu$ and $b=B_\mu \mu$ parameters, otherwise the scenario suffers from a severe $\mu/B_\mu$ problem. 
The solution occurs for $\xi\rightarrow -\pi/4$ and $\phi\neq \xi$, with a small detuning between $\phi$ and $\xi$ that controls the size of the $\mu$ term.   In this limit, the $\mathcal{R}_{u,d}$ are given to leading order by
\begin{eqnarray}
\mathcal{R}_{u,d}= \left ( \begin{array}{ccc} \frac{1}{\sqrt{3}} & \mp \frac{1}{2} \left (1+\frac{1}{\sqrt{3}} \right) & \frac{1}{2} \left (1-\frac{1}{\sqrt{3}} \right) \\  \frac{1}{\sqrt{3}} & \pm \frac{1}{2} \left (1-\frac{1}{\sqrt{3}} \right) & -\frac{1}{2} \left (1+\frac{1}{\sqrt{3}} \right) 
\\  \frac{1}{\sqrt{3}} &  \pm \frac{1}{\sqrt{3}} &  \frac{1}{\sqrt{3}} \end{array} \right ).
\label{rotationmatrices}
\end{eqnarray}
Note that the trimaximal column is associated with the light eigenstate, which is precisely the state that corresponds to the electroweak doublets $H_{u,d}$.  More precisely, the eigenvalues corresponding to this light eigenstate are $\mu\ll M$ for the case of $\mathbb{M}$, and $b\ll F$ for the case of $\mathbb{F}$.  The heavy states in this limit have equal masses $M_{\rm mess}$, that are of order $M$. The larger eigenvalues of $\mathbb{F}$ are $F_{2,3}\sim F$ (the detailed relations can be found in \cite{Everett:2016meb}).

The next step is to consider the couplings of the Higgs-messenger fields to the MSSM matter fields.  Of the variety of possibilities (see \cite{Everett:2016meb} for a classification), let us focus here on renormalizable interactions of all three generations.  This results from the specific $\mathcal{S}_3$ charge assignments summarized in Table~\ref{tab:12}.
\begin{table}[htbp]
   \centering
    \begin{tabular}{c|cccc|cccccccccc|c}
  & $\mathcal{H}_u^{(2)}$&$\mathcal{H}_u^{(1)}$ & $\mathcal{H}_d^{(2)}$& $\mathcal{H}_d^{(1)}$  & $Q_{\mathbf 2}$ &$Q_{\mathbf 1}$& $\bar u_{\mathbf 2}$ & $\bar u_{\mathbf 1 }$&$\bar d_{\mathbf 2}$& $\bar d_{\mathbf 1}$ & $L_{\mathbf{2}}$ & $L_{\mathbf{1}}$ & $\bar{e}_{\mathbf{2}}$ & $\bar{e}_{\mathbf{1}}$&$X_H$\\
    \hline
    $\mathcal{S}_3$ &$\mathbf 2 $& $\mathbf 1$& $\mathbf 2 $& $\mathbf 1 $  & $\mathbf 2 $&$\mathbf 1$  & $\mathbf 2 $&$\mathbf 1$  & $\mathbf 2 $&$\mathbf 1$ & $\mathbf 2 $&$\mathbf 1$  & $\mathbf 2 $&$\mathbf 1$ &\textbf 2\\
   \end{tabular}
   \caption{Charges for an $\st$ model of the Higgs-messenger fields and the MSSM matter fields. Here the $SU(3)$ triplet messengers and the associated $X_T$ field are not displayed for simplicity.}
   \label{tab:12}
\end{table} 
The renormalizable superpotential Yukawa couplings, for example in the up quark sector, are given as follows:
\begin{eqnarray}
W^{(u)}= y_u\big[Q_{\mathbf 2}  \bar u_{\mathbf 2}  \mathcal{H}^{(2)}_u+\beta_1Q_{\mathbf 2}  \bar u_{\mathbf 2} \mathcal{H}^{(1)}_u + \beta_2 Q_{\mathbf 2}  \bar u_{\mathbf 1}  \mathcal{H}^{(2)}_u +\beta_3 Q_{\mathbf 1}  \bar u_{\mathbf 2}  \mathcal{H}^{(2)}_u+ \beta_4 Q_{\mathbf 1}  \bar u_{\mathbf 1}  \mathcal{H}^{(1)}_u\big],
\label{wu}
\end{eqnarray}
in which the $\beta_i$ are arbitrary coefficients in the absence of further model structure, and the overall scaling $y_u$ is also a free parameter.  In the basis given by $Q=(Q_{\mathbf 2} ,Q_{\mathbf 1})^T$ and $\overline{u}=(\overline{u}_{\mathbf 2} ,\overline{u}_{\mathbf 1})^T$, these couplings can be expressed in matrix form as
\begin{eqnarray}
W^{(u)}=y_uQ^T\left( \begin{matrix} \mathcal{H}^{(2)}_{u1}&\beta_1\mathcal{H}^{(1)}_{u}&\beta_2 \mathcal{H}^{(2)}_{u2}\\ \beta_1 \mathcal{H}^{(1)}_u& \mathcal{H}^{(2)}_{u2}& \beta_2\mathcal{H}^{(2)}_{u1}\\ \beta_3\mathcal{H}^{(2)}_{u2}& \beta_3\mathcal{H}^{(2)}_{u1}&\beta_4 \mathcal{H}^{(1)}_u\end{matrix}\right)\bar u. \label{UpYukawas}
\end{eqnarray}
Analogous coupling matrices would hold in the down quark and charged lepton sectors, with the replacements $\beta_i\rightarrow \beta_{di},\beta_{ei}$~\footnote{We neglect issues of neutrino mass generation for simplicity.}. Here we will focus on the up quark sector only, and later assume that the other charged fermions follow similar patterns.  (Let us also comment that Eq.~(\ref{wu}) and its generalizations to other charged SM fermions correct a typo in the corresponding expressions for the superpotential in \cite{Everett:2016meb}, for which there was an incorrect interchange of $\beta_1$ and $\beta_2$.) 

From Eq.~(\ref{UpYukawas}) and the definition of the Higgs-messenger diagonalization matrices $\mathcal{R}_{u,d}$ as given in Eq.~(\ref{higgs_s3}) and Eq.~(\ref{rotationmatrices}), it is straightforward to see that the SM Yukawa matrix takes the form
\begin{equation}
Y_u = \frac{y_u}{\sqrt{3}} \left (\begin{array}{ccc} 1 & \beta_1 & \beta_2 \\ \beta_1 & 1 & \beta_2 \\ \beta_3 & \beta_3 & \beta_4           \end{array} \right ),
\label{eq:yu}
\end{equation}
while the messenger Yukawas are given by
\begin{equation}
Y^\prime_{u1}=y_u \left (\begin{array}{ccc} -\frac{1}{2}-\frac{1}{2\sqrt{3}} & \frac{\beta_1}{\sqrt{3}} & \;\; \frac{\beta_2}{2} - \frac{\beta_2}{2\sqrt{3}} \\  \frac{\beta_1}{\sqrt{3}} & \;\; \frac{1}{2}-\frac{1}{2\sqrt{3}} & -\frac{\beta_2}{2} - \frac{\beta_2}{2\sqrt{3}} \\ \;\; \frac{\beta_3}{2} - \frac{\beta_3}{2\sqrt{3}} & -\frac{\beta_3}{2} - \frac{\beta_3}{2\sqrt{3}} & \frac{\beta_4
}{\sqrt{3}}
\end{array} \right )
\label{eq:yu1p}
\end{equation}
\begin{equation}
\;\; Y^\prime_{u2}=y_u \left (\begin{array}{ccc} \;\; \frac{1}{2}-\frac{1}{2\sqrt{3}} & \frac{\beta_1}{\sqrt{3}} &  -\frac{\beta_2}{2} - \frac{\beta_2}{2\sqrt{3}} \\  \frac{\beta_1}{\sqrt{3}} & -\frac{1}{2}-\frac{1}{2\sqrt{3}} & \;\; \frac{\beta_2}{2} - \frac{\beta_2}{2\sqrt{3}} \\ -\frac{\beta_3}{2} - \frac{\beta_3}{2\sqrt{3}} & \;\; \frac{\beta_3}{2} - \frac{\beta_3}{2\sqrt{3}} & \frac{\beta_4
}{\sqrt{3}}
\end{array} \right ).
\label{eq:yu2p}
  \end{equation}
There are a variety of ways in which a hierarchy of fermion masses can be achieved.  For example, in \cite{Everett:2016meb}, the $\beta_i$ were all taken to be equal and set to unity, which led to two massless quark mass eigenvalues and one heavy (third generation) state.  Rather than classifying all possibilities, here we consider a specific example in which the $\beta_i$ parameters obey the following constraint:
\begin{equation}
\beta_1=1, \qquad \beta_4=\beta_2\beta_3.
\label{model1}
\end{equation}
Diagonalizing Eq.~(\ref{eq:yu}) subject to the constraint of Eq.~(\ref{model1}) results in two massless eigenvalues, and one nonzero eigenvalue,
\begin{equation}
y_t= y_u \left (\sqrt{2+\beta_2^2}\sqrt{2+\beta_3^2} \right)/\sqrt{3}.
\end{equation}
Furthermore, as $y_u$ and $\beta_{2,3}$ are all arbitrary parameters, we will further consider the limit in which $\beta_{2,3}$ are taken to be very large, while $y_u$ is simultaneously taken to be very small, such that $y_t$ as given above remains fixed.  We see from Eq.~(\ref{eq:yu}) that in this limit, we have
\begin{equation}
Y_u = \frac{y_t}{\sqrt{2+\beta_2^2}\sqrt{2+\beta_3^2}} \left (\begin{array}{ccc} 1 & 1 & \beta_2 \\ 1 & 1 & \beta_2 \\ \beta_3 & \beta_3 & \beta_2\beta_3           \end{array} \right ) \stackrel{ \beta_{i}\gg 1} {\longrightarrow} \; {\rm Diag}(0,0,y_t).
\label{yulimit}
\end{equation}
Again imposing Eq.~(\ref{model1}), it is easy to see from Eqs.~(\ref{eq:yu1p}) and (\ref{eq:yu2p}) that in the limit of large $\beta_{2,3}$ and fixed $y_t$, the messenger Yukawa couplings $Y^\prime_{u1}$ and $Y^\prime_{u2}$ also reduce to this form:
\begin{equation}
Y^\prime_{u1} \stackrel{ \beta_{i}\gg 1} {\longrightarrow} \; {\rm Diag}(0,0,y_t), \qquad Y^\prime_{u2} \stackrel{ \beta_{i}\gg 1} {\longrightarrow} \; {\rm Diag}(0,0,y_t).
\label{messyulimit}
\end{equation}
The feature that the messenger Yukawas and the SM Yukawas are flavor-diagonal and have nonzero $33$ entries only in this limit is a consequence of the fact that the $\mathcal{S}_3$ singlet contributions have been taken to dominate over the $\mathcal{S}_3$ doublet contributions.  Indeed, an inspection of Eq.~(\ref{wu}) shows that in the regime in which Eq.~(\ref{model1}) is satisfied, and $\beta_{2,3} \rightarrow \infty$ with $y_u\rightarrow 0$ such that $y_t$ is fixed, the only term that remains is $Q_{\mathbf 1}  \bar u_{\mathbf 1}  \mathcal{H}^{(1)}_u$.   This clearly requires an additional symmetry that allows for the $\beta_4$ term to dominate while maintaining consistency with Eq.~(\ref{wh}) (this is an additional model-building complication, but not an insurmountable one \footnote{For example, one possibility is to impose a $Z_7$ symmetry, with charges $Q_{Q_{\mathbf 2}} = 1$, $Q_{\bar u_{\mathbf 2}} = 1$,  $Q_{Q_{\mathbf 1}} =2$, $Q_{\bar u_{\mathbf 1}} = 2$, $Q_{\mathcal{H}^{(1)}_u} =3$, $Q_{\mathcal{H}^{(2)}_u}=3$, $Q_{X_H}=1$, and to introduce a flavon field $\phi$ which has $Q_\phi=1$.  This suppresses the terms in the upper $2\times2$ block of $Y_u$ such that they require two insertions of the flavon field, while the other off-diagonal terms are generated with one flavon insertion.  Please note, however, that the condition that $\beta_1$=1 requires additional symmetries to avoid fine-tuning.}).

To reiterate, we have seen that we can simultaneously obtain sizable third generation SM and messenger Yukawa couplings, and messenger Yukawa couplings that are flavor-diagonal in the limit that the $\mathcal{S}_3$ singlet terms dominate over the interactions involving $\mathcal{S}_3$ doublets.  Hence, to leading order in this parameter regime, this case is equivalent to one in which there are only third generation superpotential Yukawa couplings at the renormalizable level, and all other interactions result from higher-dimensional operators \cite{Everett:2016meb}.  
However, despite the equivalency of the two cases at leading order, they can be very different at subleading order. More precisely, the path to this limiting case is highly dependent not only on the $\mathcal{S}_3$ charges of the SM fields, but also the breaking of the additional symmetries that are required to reach this parameter regime.  For example, with the $\mathcal{S}_3$ assignment in Table~\ref{tab:12}  and restricting to renormalizable couplings, it is straightforward to see that depending on how the massless eigenvalues are lifted by perturbing about Eq.~(\ref{model1}), the subleading contributions to the messenger Yukawa interactions in this limit arise in either the $23$ or $13$ sectors.  In contrast, if Eq.~(\ref{model1}) is maintained, and the first and second generations acquire masses through nonrenormalizable operators, the structure of the subleading corrections to the messenger Yukawa couplings are highly dependent on the model-building details.  Both situations have important implications not only for the masses of the superpartners, but also for questions of flavor violation, which is a critically important issue for flavored gauge mediation models.  A detailed discussion of these possibilities and their phenomenological implications is the subject of future work \cite{fgmpaper3}.

With the simple forms of the SM Yukawa couplings and the messenger Yukawa couplings, as given in Eqs. (\ref{yulimit}) and (\ref{messyulimit}), respectively (as well as their analogues in the down quark and charged lepton sectors), the corrections to the soft supersymmetry breaking terms due to the messenger interactions are easily calculated  by standard procedures.  These procedures have been given in previous literature (see e.g.~\cite{Abdullah:2012tq,Evans:2013kxa,Jelinski:2015voa}), and summarized for this set of scenarios in \cite{Everett:2016meb}.  

With the assumption that the doublet and triplet messengers both result in the same quantity $\Lambda = F_{2,3}/M_{\rm mess}\sim F/M$, which we will always assume to be the case in this paper, the leading order nonvanishing corrections to the soft supersymmetry breaking parameters are given by
\begin{align}
\label{susy0}
\left(\delta m^2_{\tilde u}\right)_{33}&=\frac{\Lambda^2y_t^2}{(4\pi)^4}\left(-\frac{52}{15}g_1^2 - 12g_2^2 - \frac{64}3g_3^2+ 8y_b^2 + 72y_t^2\right),\\
\left(\delta m^2_{\tilde d}\right)_{33}&=\frac{\Lambda^2y_b^2}{(4\pi)^4}\left(-\frac{28}{15}g_1^2 - 12g_2^2 - \frac{64}3g_3^2+ 72y_b^2 + 8y_t^2+16y_\tau^2\right),\\
\left(\delta m^2_{\tilde e}\right)_{33}&=\frac{\Lambda^2y_\tau^2}{(4\pi)^4}\left(-\frac{36}{5}g_1^2 - 12g_2^2 + 48y_b^2 + 40y_\tau^2\right),\\
\left(\delta m^2_{\tilde Q}\right)_{33}&=\frac 1 2 (\delta m^2_{\tilde u,33} + \delta m^2_{\tilde d,33}),\\
\left(\delta m^2_{\tilde L}\right)_{33}&= \frac 1 2 \delta m^2_{\tilde 3,33},\\
\delta m^2_{\tilde H_u}&=-\frac{\Lambda^2}{(4\pi)^4}\left(18 y_b^4+6y_b^2y_t^2\right),
\qquad 
\delta m^2_{\tilde H_d}
=-\frac{\Lambda^2}{(4\pi)^4}\left(18 y_b^4+6y_b^2y_t^2+12y_\tau^4\right),\\
\left(\tilde{A}_{u}\right)_{33} &=-\frac{2y_t\Lambda}{(4\pi)^2} (y_b^2 + 3y_t^2)\equiv \tilde{A}_t, \;\;\; 
\left(\tilde{A}_{d}\right)_{33} =-\frac{2y_b\Lambda}{(4\pi)^2} (y_t^2 + 3y_b^2),\;\;\;
\left(\tilde{A}_{e}\right)_{33} =-\frac{6y_\tau^3\Lambda}{(4\pi)^2}.
\label{susy1}
\end{align}
All nontrivial corrections in the squark and slepton sectors thus involve only third generation fields
\footnote{Here we note that the $\tilde{A}$ notation denotes the fact that the trilinear scalar couplings in the Lagrangian are of the form $A_{ijk}\phi_i\phi_j\phi_k$, for scalar fields $\phi_{i,j,k}$ (these are not family indices, but general field labels).}.

\section{Results}
In this section, we present a detailed numerical exploration of this scenario, as encoded by the soft supersymmetry breaking terms of Eqs.~(\ref{susy0})--(\ref{susy1}).  The model parameters are $M_{\rm mess}$, $\Lambda$, $\tan\beta$, and the sign of $\mu$ ($\text{sgn}(\mu)$), where we have followed standard procedures and replaced $\mu$ and $b$ by  $\tan\beta$, $\text{sgn}(\mu)$, and the $Z$ boson mass.  Here we will always set $\text{sgn}(\mu)=1$.  Note that in comparison to the case of minimal gauge mediation and flavored gauge mediation models based on Abelian symmetries, the number of vectorlike messenger pairs in this scenario is always fixed to be $N_5=2$, which is the smallest number that allows for separate fine-tuning of the $\mu$ and $b$ parameters.  As a result, the model studied here has one fewer discrete parameter than these other scenarios.  The renormalization group equations are evaluated using SoftSUSY 4.1.4 \cite{Allanach:2001kg}. 

In \cite{Everett:2016meb}, a preliminary analysis of this scenario was carried out in the context of taking the third generation MSSM matter fields to be inert to the $\mathcal{S}_3$ symmetry, with the primary goal of comparing this case to the case of minimal (flavor-independent) gauge mediation for a fixed value of $\tan\beta$ (of $\tan\beta=10$).  Our main purpose here is to provide a systematic analysis of the superpartner mass spectra, highlighting the dependence on $\tan\beta$ subject to the Higgs mass constraint. 

\begin{figure}[htbp]
   \centering
   \includegraphics[width = 0.45\textwidth]{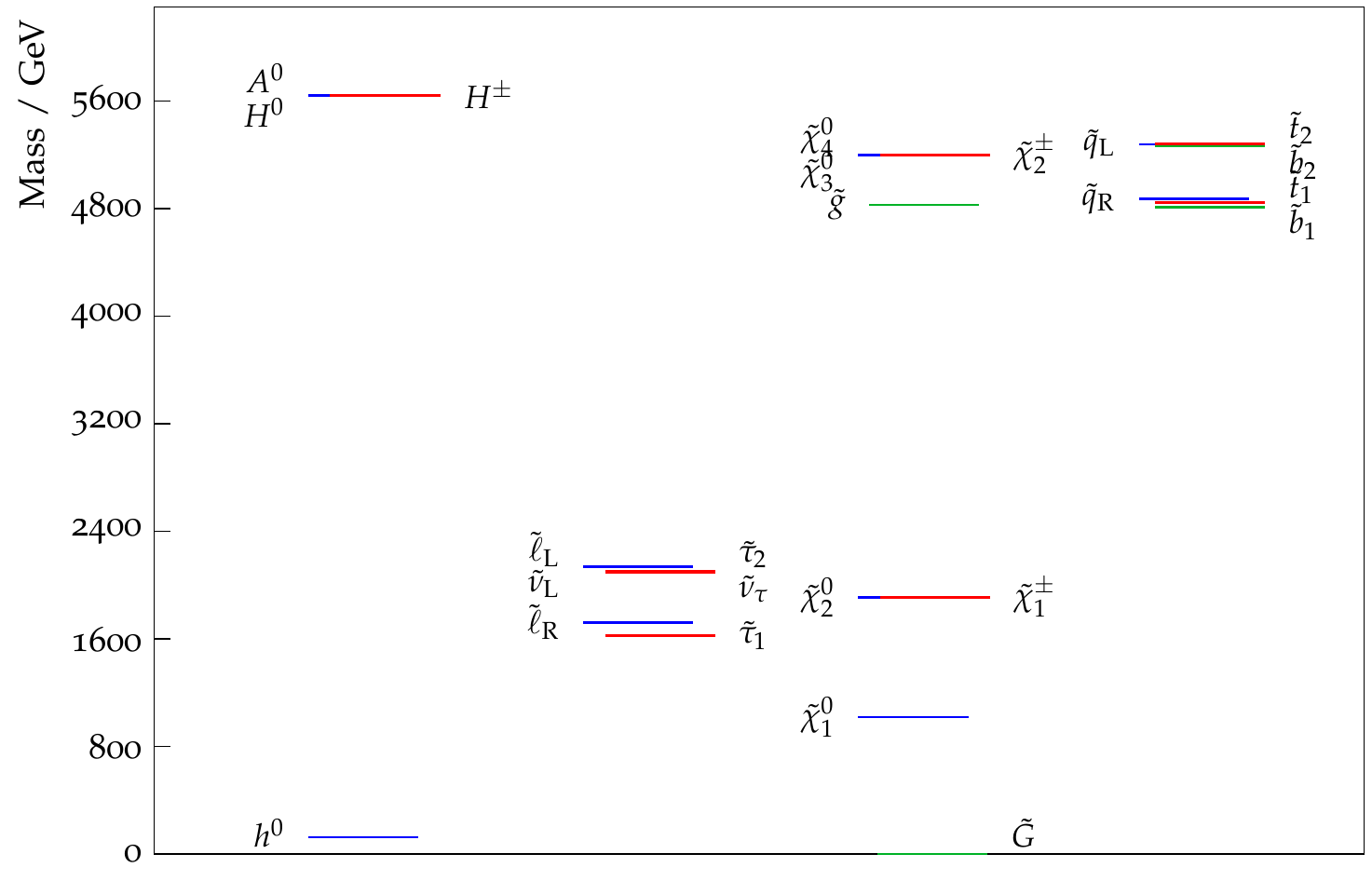} 
   \includegraphics[width = 0.45\textwidth]{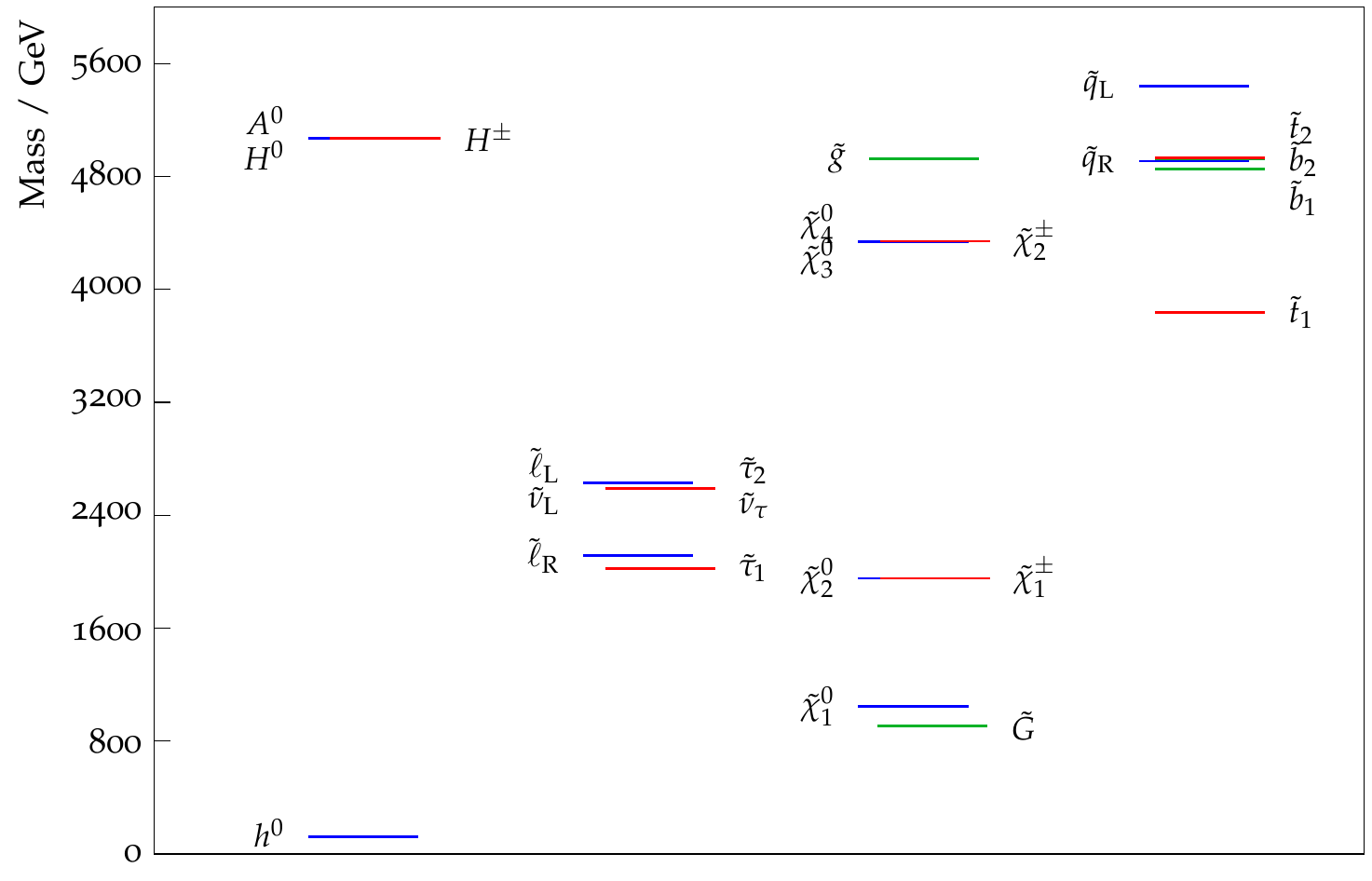}
   \centering
      \caption{The mass spectrum for $M_{\text{Mess}}= 1\times10^{12}$ GeV (left panel) and $M_{\text{Mess}}= 1\times10^{16}$ GeV (right panel), both with  $\tan\beta=10$.  In each case, $\Lambda$ is fixed by the Higgs mass constraint.}
   \label{fig:Case1Example1}
\end{figure}

For concreteness, and to connect with the results of our previous work, we begin with the example shown in the left panel of Figure~\ref{fig:Case1Example1}, for which the messenger scale is $M_{\rm mess} = 10^{12}$ GeV and $\tan\beta=10$; once these two parameters are fixed, the value of $\Lambda$ is thus set by the requirement that  $m_h\simeq 125$ GeV.  
This example shows a characteristic pattern also seen in the intermediate scale examples of \cite{Everett:2016meb}, for which the heavy Higgs particles are between $5-6$ TeV, the gluino is approximately $5$ TeV,  and the squarks fall into two groupings, a lighter set that is close in mass to the gluino, and a heavier set that is similar in mass to the heavy charginos and neutralinos.  The sleptons are close in mass to the lightest chargino and the second-lightest neutralino, which are nearly mass degenerate, and the next-to-lightest superpartner (NLSP) is the lightest neutralino. The needed values of $\mu$ and $b/\mu$ are in the $5-6$ TeV range.  With an intermediate messenger scale, the gluino is lighter than the masses of the sparticles controlled by $\mu$.  
In the right panel of Figure~\ref{fig:Case1Example1}, we show an example with $M_{\rm mess}= 10^{16}$ GeV and $\tan\beta=10$.  Due to increased renormalization group running effects, the $\mu$ and $b/\mu$ terms are lighter than in the previous case, which leads to lighter masses for the heavy Higgs states. The gluino is now heavier than the heavy charginos and neutralinos, which have masses controlled by $\mu$.  There are also larger stop mixing effects, with the lighter stop mass just below 4 TeV, and a heavier gravitino due to the high messenger scale.

\begin{figure}[htbp]
\begin{subfigure}
   \centering
   \includegraphics[width = 0.45\textwidth]{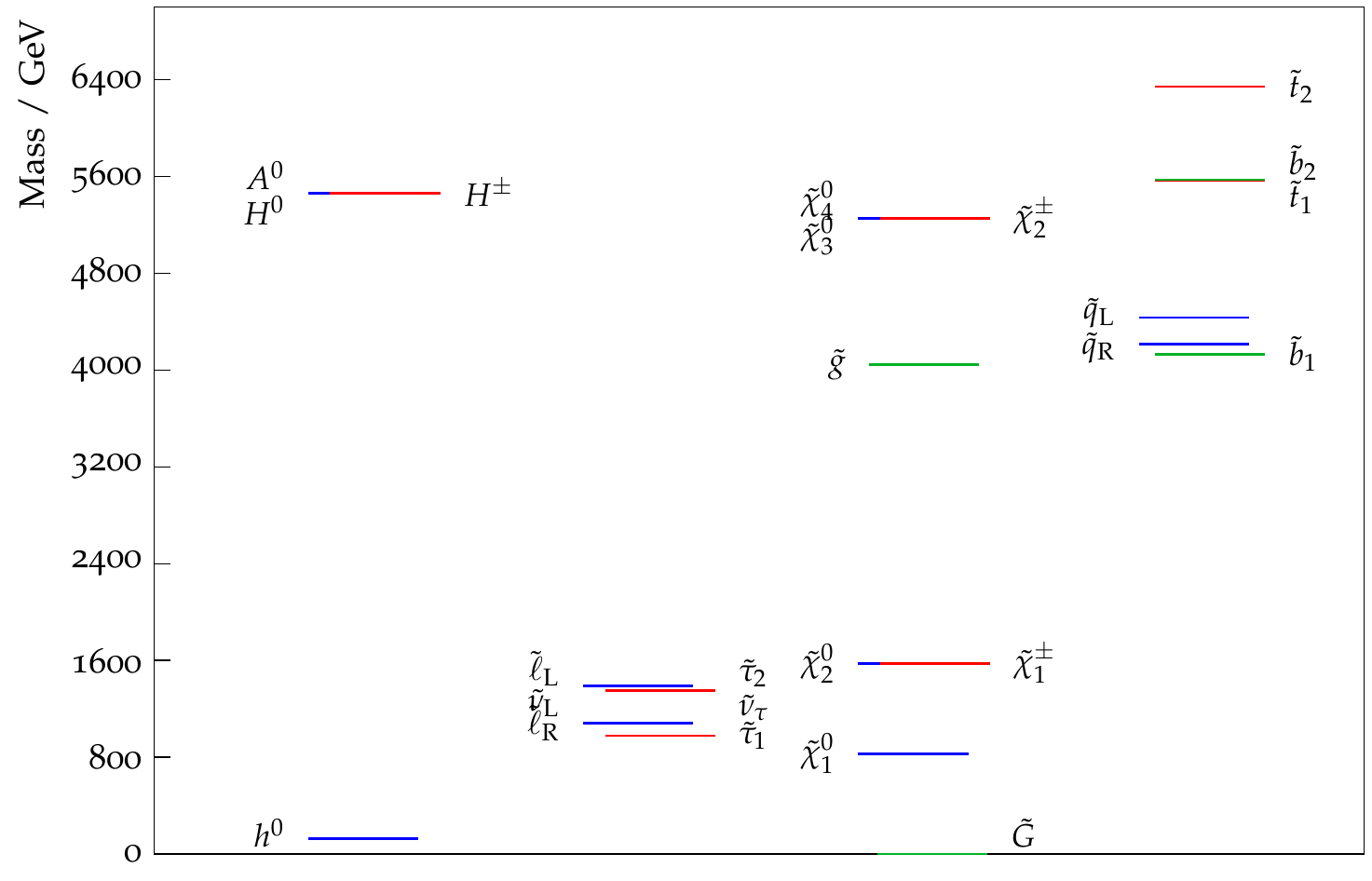} 
   \end{subfigure}
   \begin{subfigure}
   \centering
   \includegraphics[width = 0.45\textwidth]{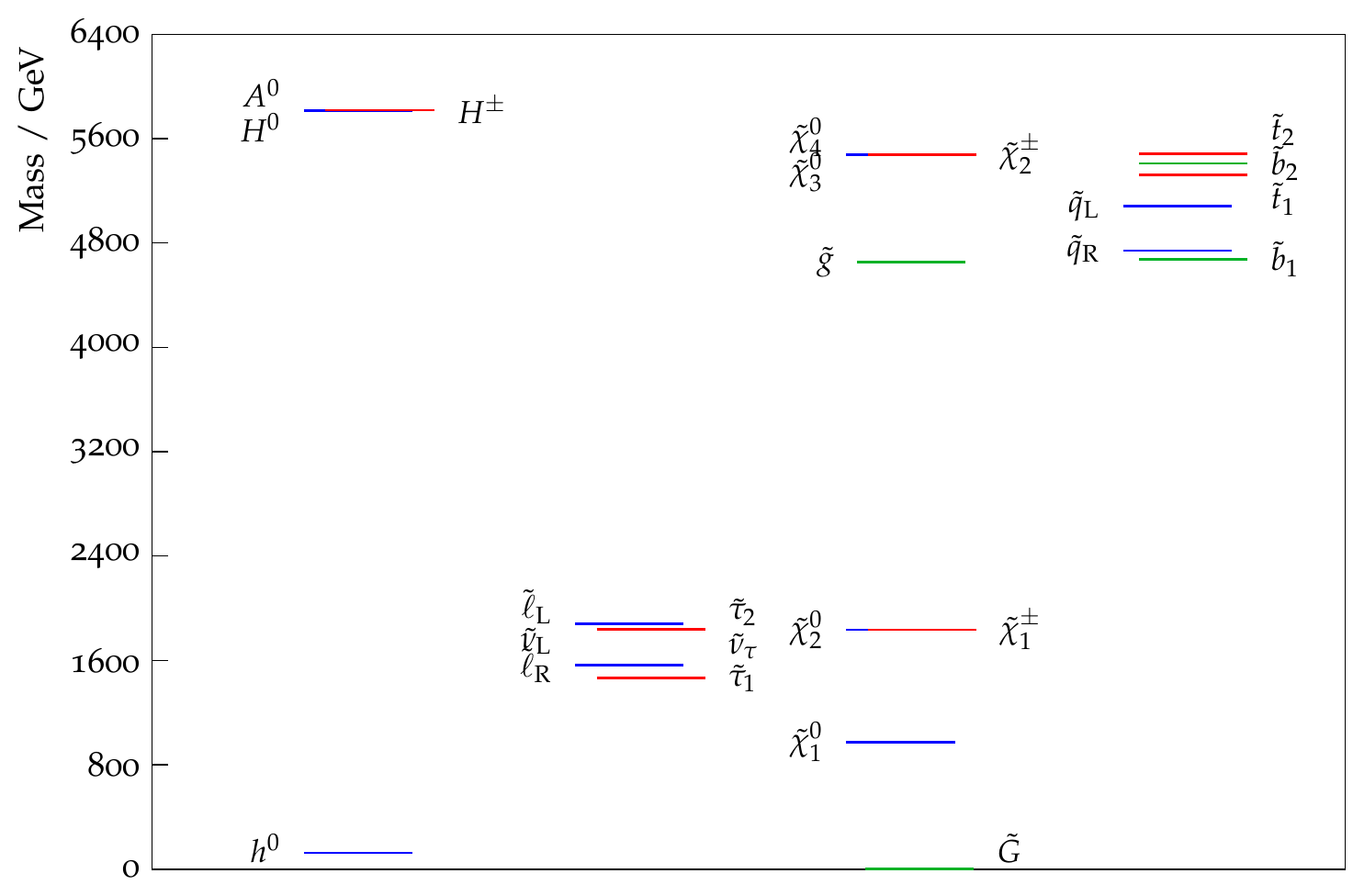}
   \end{subfigure}
   \centering
      \caption{The mass spectrum for $M_{\text{Mess}}= 1\times10^{6}$ GeV (left panel) and $M_{\text{Mess}}= 1\times10^{10}$ GeV (right panel), both with  $\tan\beta=10$.  In each case, $\Lambda$ is fixed by the Higgs mass constraint.}
   \label{fig:Case1Example2}
\end{figure}

For smaller values of the messenger scale,  the spectra are generally heavier for a fixed $\tan\beta$, as the stop mixing is smaller, requiring an increase in the value of $\Lambda$ to obtain the same value of the light Higgs mass.  This is seen in Figure~\ref{fig:Case1Example2}, which shows points with $M_{\rm mess} = 10^6$ GeV (left panel) and $M_{\rm mess}=10^{10}$ GeV (right panel), both with $\tan\beta = 10$. We note the lighter sleptons, lighter gluino mass, and greater squark mass splitting for low messenger scales.

\begin{figure}[htbp]
\begin{subfigure}
   \centering
   \includegraphics[width = 0.45\textwidth]{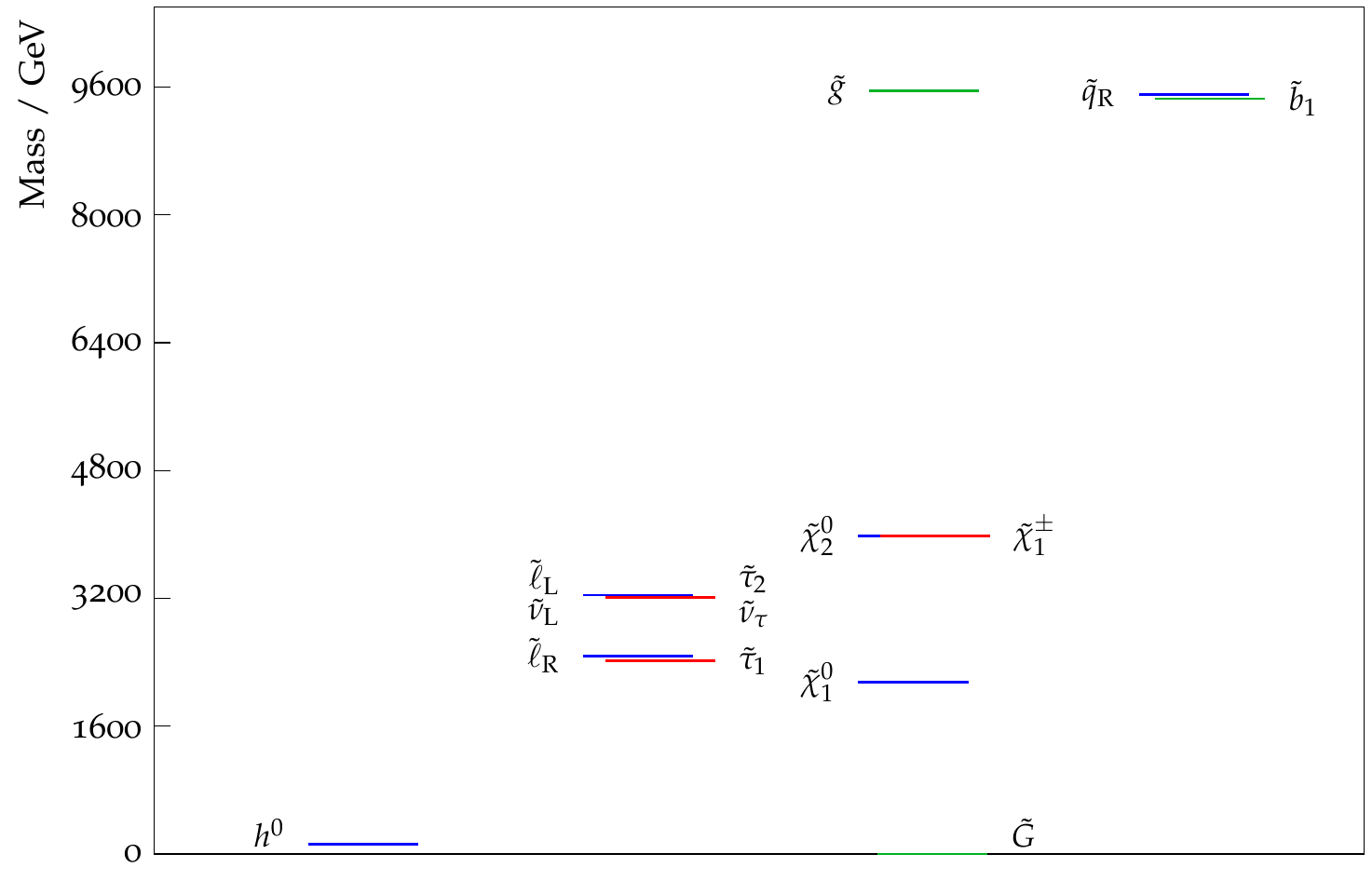} 
   \end{subfigure}
   \begin{subfigure}
   \centering
   \includegraphics[width = 0.45\textwidth]{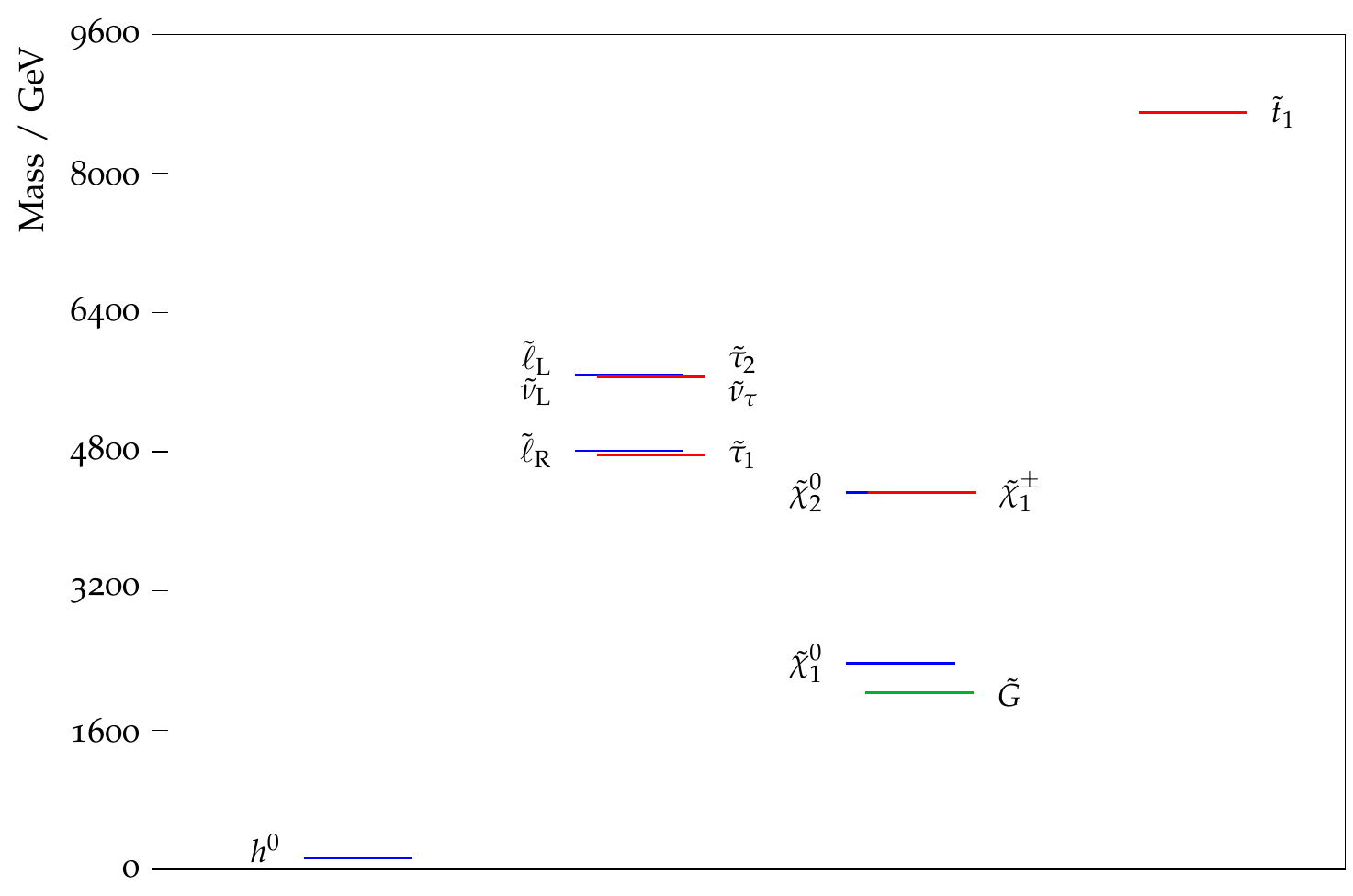}
   \end{subfigure}
   \centering
      \caption{The mass spectrum for $M_{\text{Mess}}= 1\times10^{6}$ GeV (left panel) and $M_{\text{Mess}}= 1\times10^{12}$ GeV (right panel), both with  $\tan\beta=5$.  In each case, $\Lambda$ is fixed by the Higgs mass constraint.}
   \label{fig:Case1Example3}
\end{figure}

Let us now consider smaller values of $\tan\beta$.  In Figure~\ref{fig:Case1Example3}, we show a low messenger scale example with $M_{\rm mess}=10^6$ GeV (left panel) and a high messenger scale example with $M_{\rm mess}=10^{12}$ GeV (right panel), both now with $\tan\beta=5$. The sparticle masses are expected to be heavier for smaller $\tan\beta$, since the tree-level contribution to the light Higgs mass has decreased, requiring even larger radiative corrections to boost the light Higgs mass to its experimentally measured value.  As a result, the mass spectra in these cases are highly split, and even the lighter sparticles have masses at and above 2 TeV.  For higher values of the messenger scale, this splitting is amplified, as a larger value of $\Lambda$ is needed to compensate for smaller $A$ terms at  low energies.   These features are clearly exhibited in Figure~\ref{fig:Case1Example3}, where we note that in each case, the heavy Higgs bosons and many of the squark masses are in the 10 TeV range, and thus out of the range shown in the figure.

\begin{figure}[htbp]
\begin{subfigure}
   \centering
   \includegraphics[width = 0.45\textwidth]{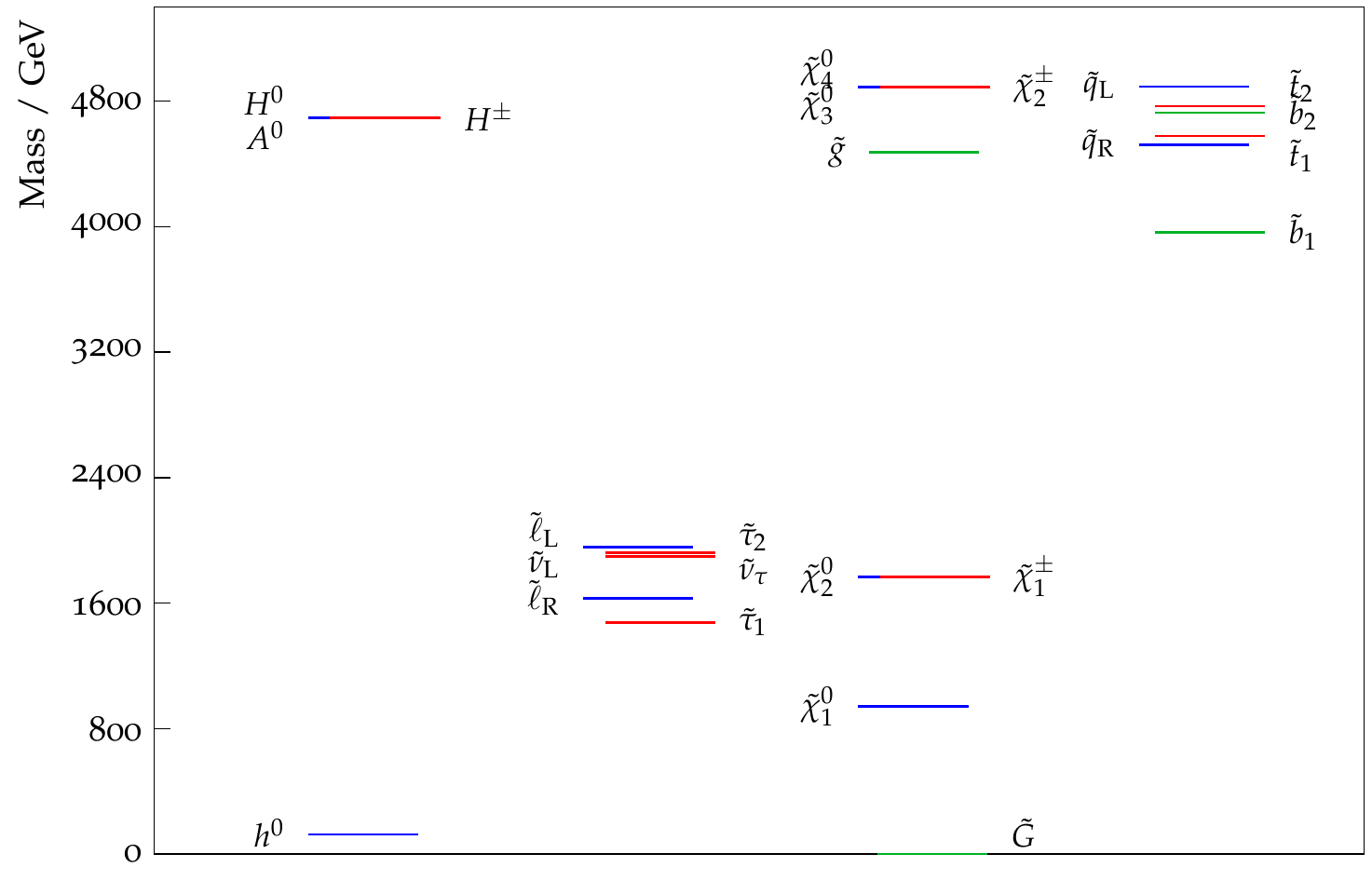} 
   \end{subfigure}
   \begin{subfigure}
   \centering
   \includegraphics[width = 0.45\textwidth]{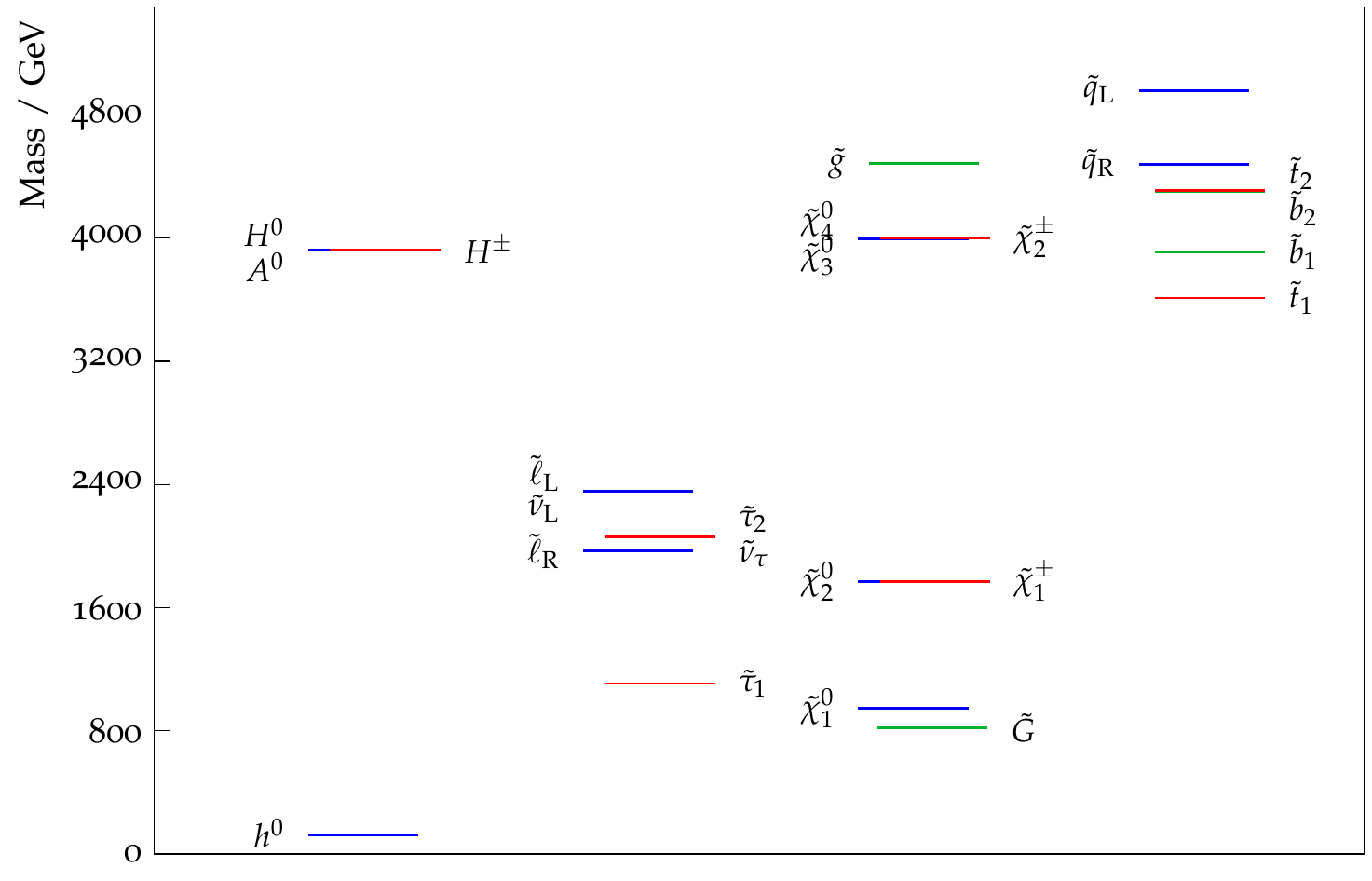}
   \end{subfigure}
   \centering
      \caption{The mass spectrum for $M_{\text{Mess}}= 1\times10^{12}$ GeV (left panel) and $M_{\text{Mess}}= 1\times10^{16}$ GeV (right panel), both with  $\tan\beta=40$.  In each case, $\Lambda$ is fixed by the Higgs mass constraint.}
   \label{fig:Case1Example4}
\end{figure}

We now consider the limit of large $\tan\beta$, for which the effects of the bottom and tau Yukawa couplings are more significant than in the lower $\tan\beta$ regime.  In Figure~\ref{fig:Case1Example4}, we show spectra with $M_{\rm mess}=10^{12}$ GeV (left panel) and $M_{\rm mess}=10^{16}$ GeV (right panel), both with $\tan\beta=40$.  In comparison to the analogous $\tan\beta = 10$ cases as shown in Figure~\ref{fig:Case1Example1}, the higher $\tan\beta$ spectra are compressed, with the heaviest superpartner masses in the 5 TeV range.  The two cases again differ in their values of $\mu$ and $b/\mu$, with the high messenger scale example again displaying smaller values for these quantities, such that the gluino is heavier than the heavy charginos and neutralinos, and the heavy Higgs particles are lighter than their counterparts in the intermediate scale case.  We also see a greater splitting of the squark masses in the high messenger scale case.
\begin{figure}[htbp]
\begin{subfigure}
   \centering
   \includegraphics[width = 0.45\textwidth]{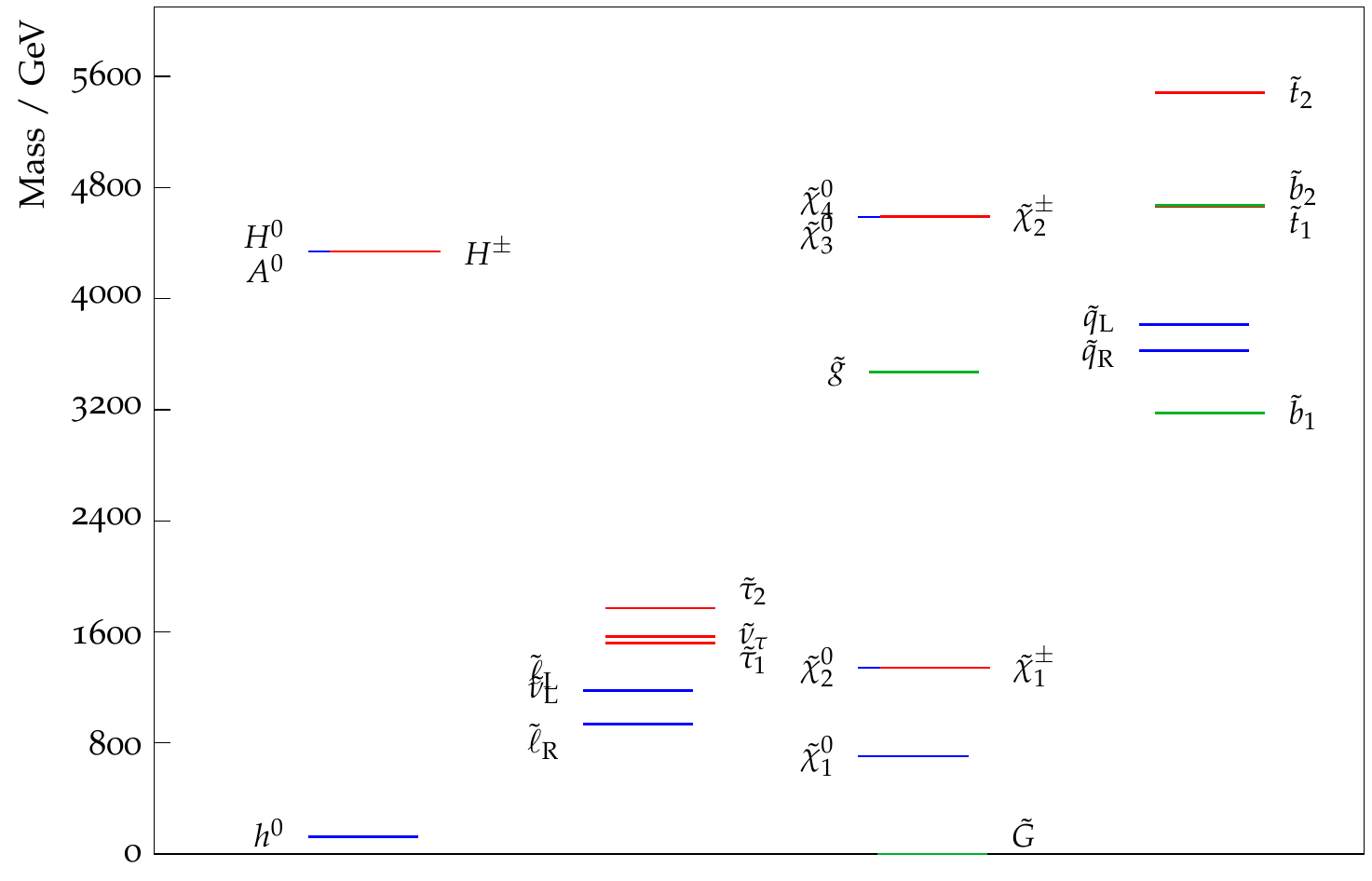} 
   \end{subfigure}
   \begin{subfigure}
   \centering
   \includegraphics[width = 0.45\textwidth]{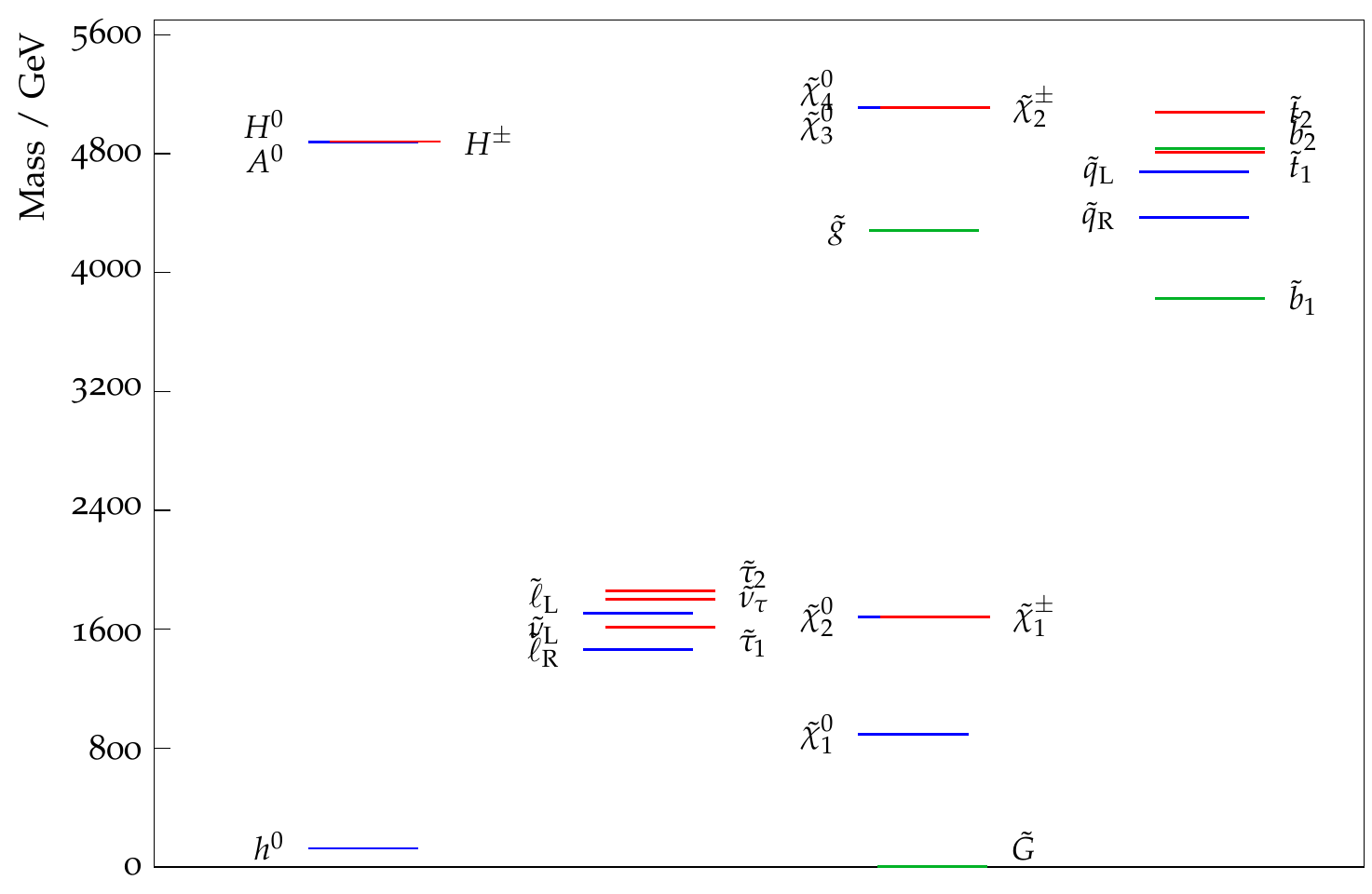}
   \end{subfigure}
   \centering
      \caption{The mass spectrum for $M_{\text{Mess}}= 1\times10^{6}$ GeV (left panel) and $M_{\text{Mess}}= 1\times10^{10}$ GeV (right panel), both with  $\tan\beta=40$.  In each case, $\Lambda$ is fixed by the Higgs mass constraint.}
   \label{fig:Case1Example5}
\end{figure}
These similarities continue in Figure~\ref{fig:Case1Example5}, which is the analogous set of mass spectra to Figure~\ref{fig:Case1Example2}, with $M_{\rm mess}=10^6$ GeV (left panel) and $M_{\rm mess}=10^{10}$ GeV, but now with $\tan\beta=40$.  As expected, the superpartner mass spectrum is again compressed compared to lighter values of $\tan\beta$, though slightly less so than what occurs for higher messenger scale values.  For the low messenger scale example, there is a relatively light gluino and light bottom squark, with masses in the 3 TeV range.  The NLSP remains the lightest neutralino, though the lighter stau mass continues to approach the NLSP mass as the messenger scale increases, due to the larger value of the tau Yukawa coupling in the large $\tan\beta$ regime.

\begin{figure}[htbp]
   \centering
   \includegraphics[width = 0.65\textwidth]{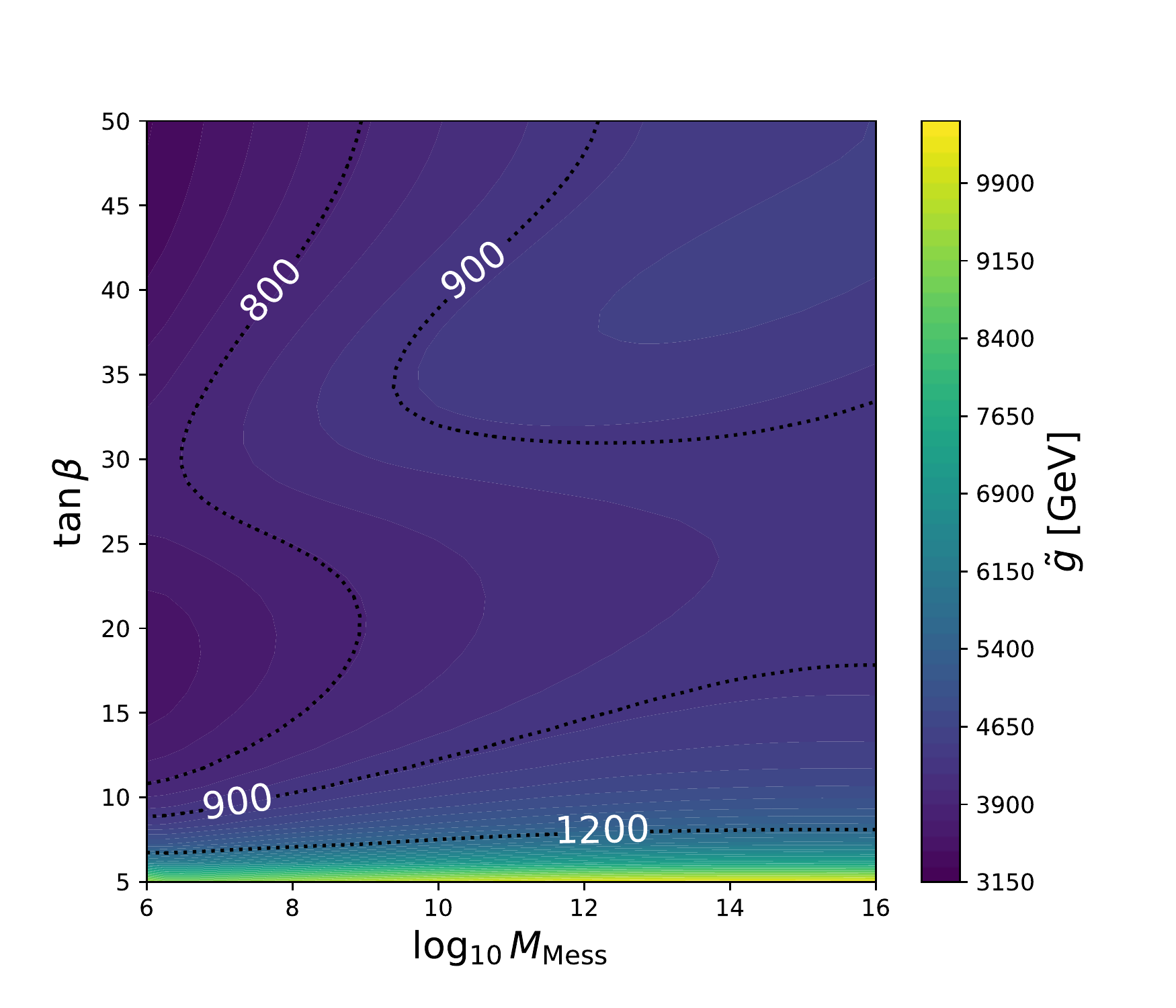} 
   \centering
      \caption{The range of gluino masses in this scenario, as displayed in the $\log M_{\rm mess}-\tan\beta$ plane. The dotted contours denote the lightest neutralino mass.}
   \label{fig:Case1Example6}
\end{figure}

These representative examples of the superpartner mass spectra demonstrate that as $\tan\beta$ is varied, there is a range of cases in the low $\tan\beta$ regime that have very heavy sparticle masses, but otherwise we have a range of superpartner masses that tend to be at most in the $5-6$ TeV range.  This behavior in fact is quite robust. In Figure~\ref{fig:Case1Example6}, the range of gluino masses in the $\log(M_{\rm mess})-\tan\beta$ plane is shown, with the NLSP (lightest neutralino) mass given by the dotted contours.  There is a tiny sliver of parameter space, corresponding to the lowest allowed values of $\tan\beta$, for which the gluino mass (and $\Lambda$) increases substantially, and increases as the messenger scale increases.  Otherwise, the parameter regime is dominated by gluino masses within the $4-5$ TeV range, as seen repeatedly in the example spectra shown here.

\begin{figure}[htbp]
\begin{subfigure}
   \centering
   \includegraphics[width = 0.45\textwidth]{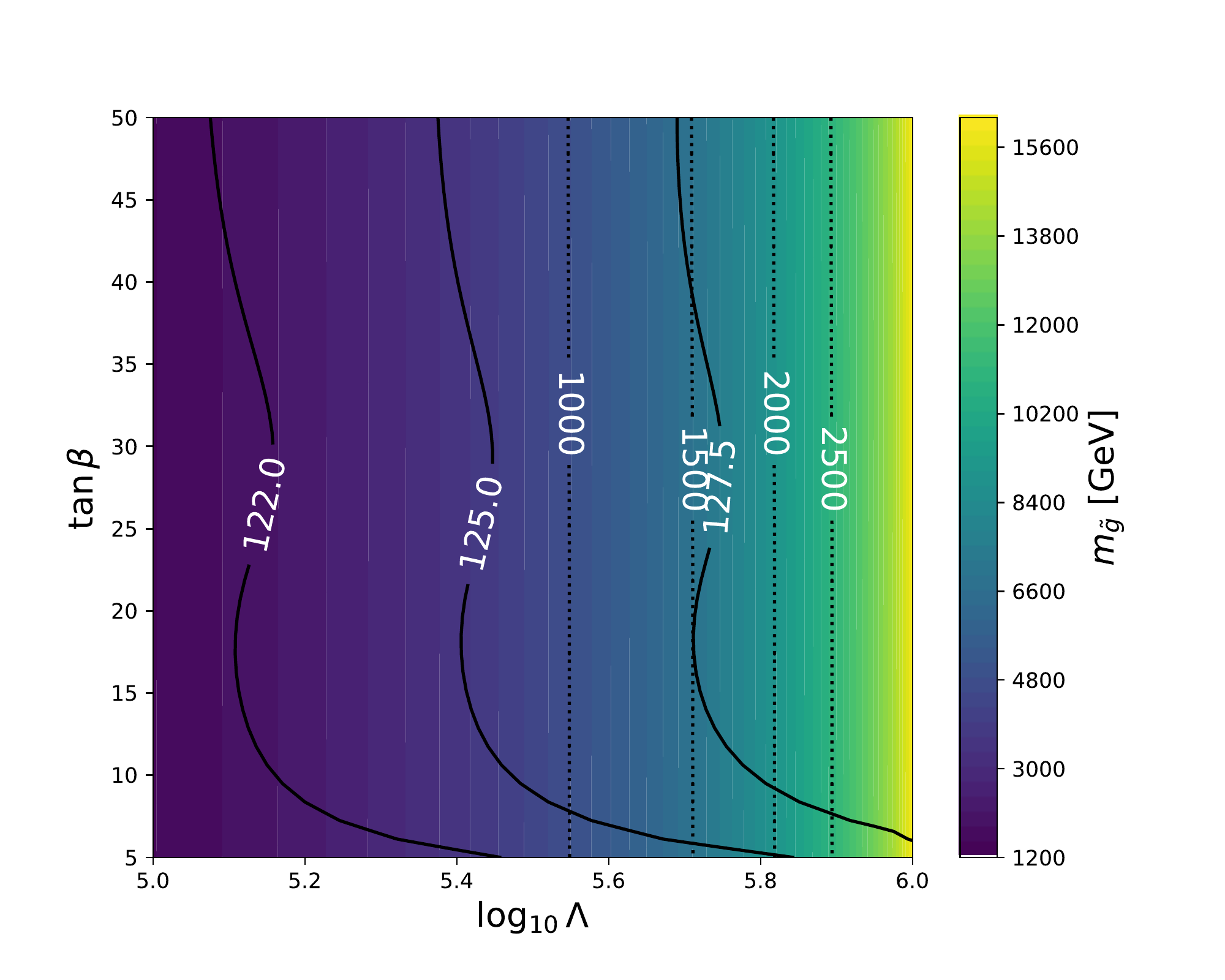} 
    \end{subfigure}
   \begin{subfigure}
   \centering
    \includegraphics[width = 0.45\textwidth]{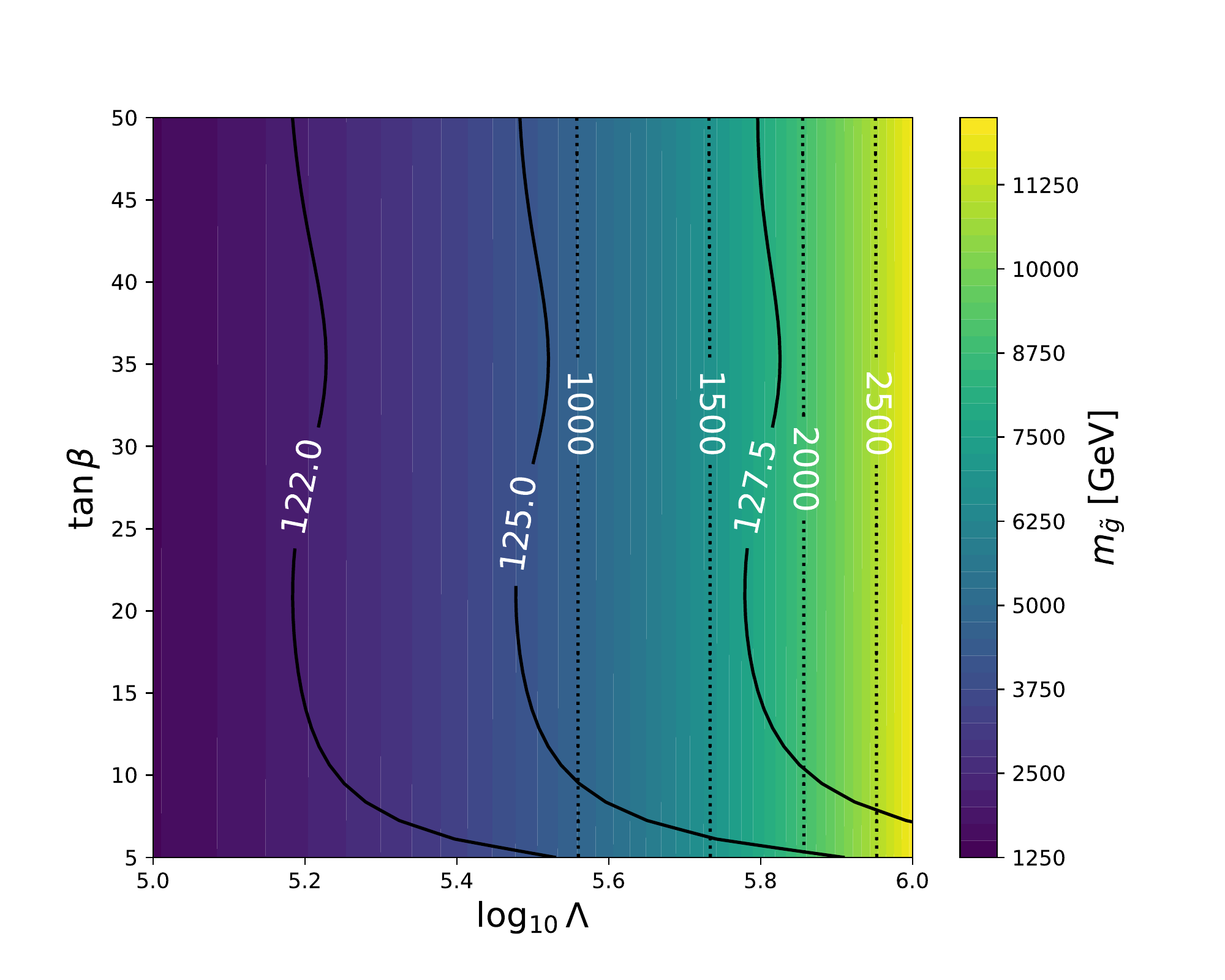} 
    \end{subfigure}
    \centering
      \caption{The Higgs mass (solid contours) and gluino masses (color shading) in this scenario, as displayed in the $\log_{10}\Lambda-\tan\beta$ plane for $M_{\rm mess}=10^6$ GeV (left panel) and $M_{\rm mess}=10^{10}$ GeV (right panel).  The dotted contours show the lightest neutralino mass.}
   \label{fig:contourlambdatanbeta10}
\end{figure}
It is also illuminating to explore the Higgs mass prediction and its correlation with the gluino mass within the parameter space of this model.  In Figure~\ref{fig:contourlambdatanbeta10}, we show the $\log_{10}(\Lambda)-\tan\beta$ plane for two fixed values of the messenger scale: $M_{\rm mess}=10^{6}$ GeV (left panel) and $M_{\rm mess}=10^{10}$ GeV (right panel).  The Higgs mass contours are given by solid lines, the gluino mass is labeled by color shading, and the dotted contours represent the lightest neutralino (NLSP) mass.  Here we see the correlation in this model between the needed values of the gluino mass and the Higgs mass constraint, with gluino masses in the range of 4 TeV or greater for the experimentally preferred range of the lightest Higgs mass.  The NLSP mass in the allowed parameter region is correspondingly lighter for lower messenger scale values than for higher values.  The general features of the $M_{\rm mess}=10^{10}$ GeV case persist for higher values of the messenger scale (not displayed here for simplicity).
\begin{figure}[htbp]
\begin{subfigure}
   \centering
   \includegraphics[width = 0.45\textwidth]{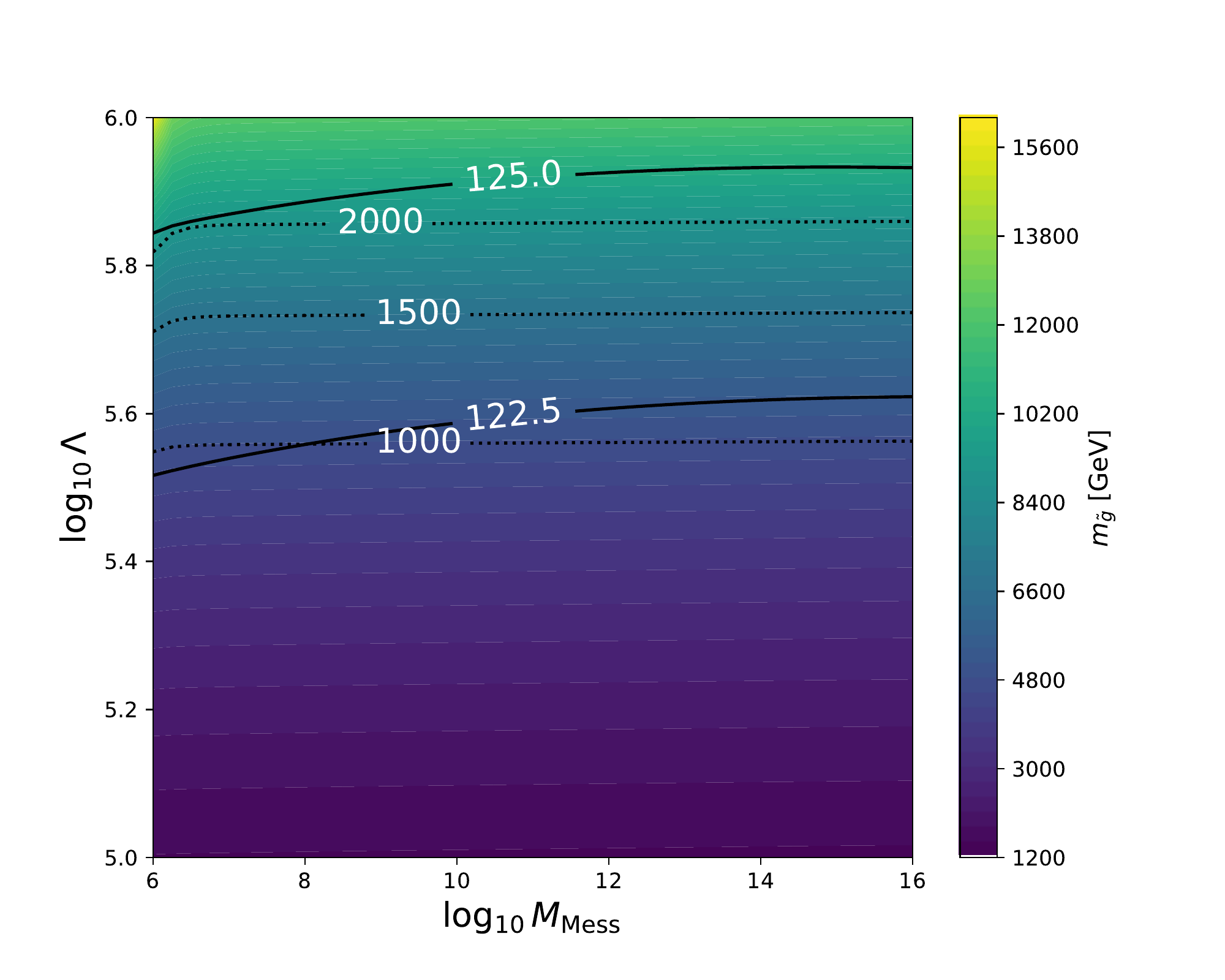} 
    \end{subfigure}
   \begin{subfigure}
   \centering
    \includegraphics[width = 0.45\textwidth]{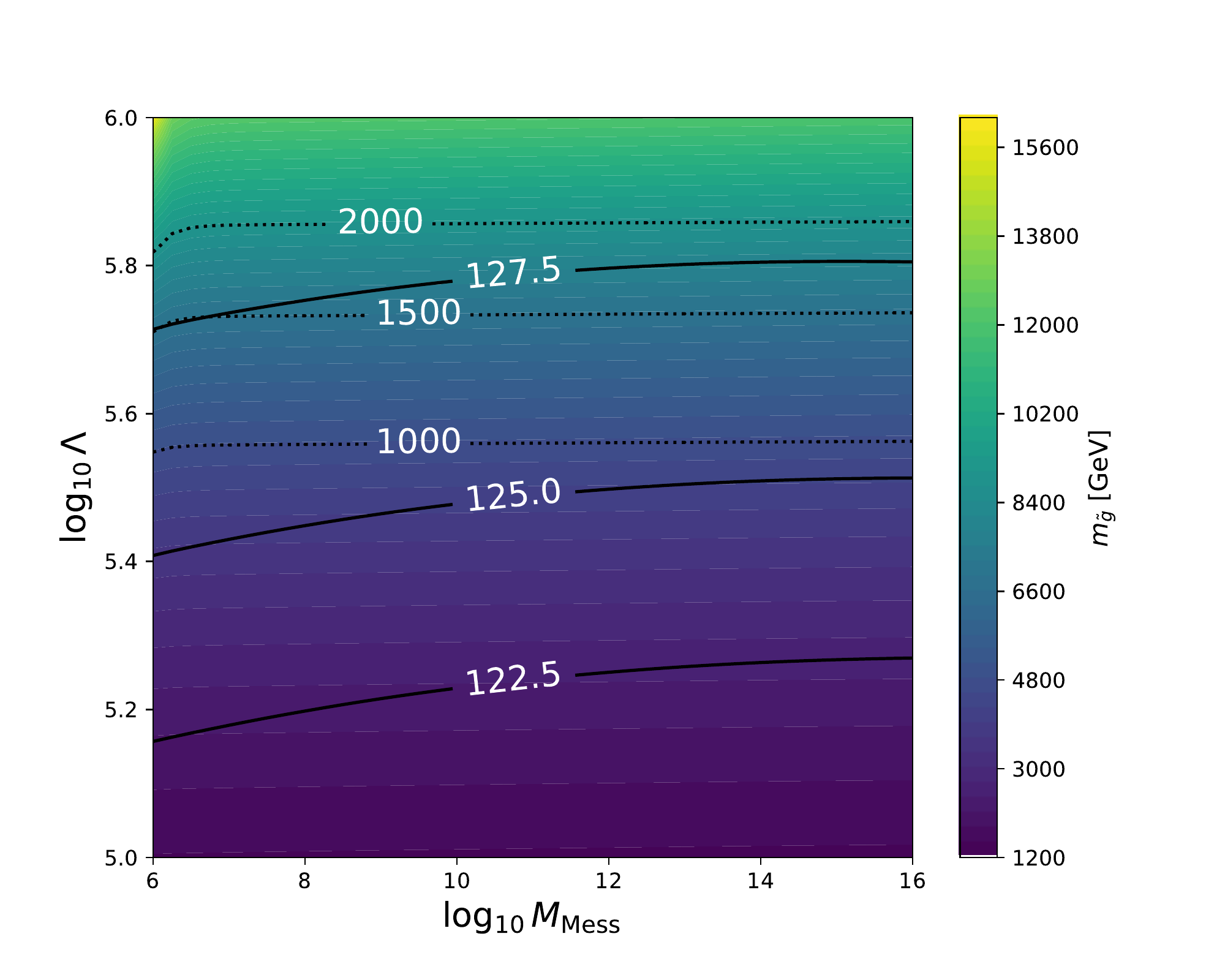} 
    \end{subfigure}
    \centering
      \caption{The Higgs mass (solid contours) and gluino masses (color shading) in this scenario, as displayed in the $\log_{10}\Lambda-\log_{10}M_{\rm mess}$ plane for $\tan\beta=5$ (left panel) and $\tan\beta=20$ (right panel).  The dotted contours show the lightest neutralino mass.}
   \label{fig:contourmmesslambda10}
\end{figure}

In Figure~\ref{fig:contourmmesslambda10}, the Higgs and gluino mass predictions are displayed in the $\log_{10}M_{\rm mess}-\log_{10}\Lambda$ plane for $\tan\beta =5$ (left panel) and $\tan\beta = 20$ (right panel).  The contour labeling is identical to Figure~\ref{fig:contourlambdatanbeta10}, with the Higgs mass given by solid contours, the NLSP mass by dotted contours, and the gluino mass as color shading;  note the different scale of the color shading compared to the previous figure.  Here we see the differences in the gluino mass predictions between the low $\tan\beta$ regime and the intermediate to high $\tan\beta$ regimes, with significantly heavier gluino masses needed for very low $\tan\beta$, and the $<5$ TeV range for intermediate to high $\tan\beta$.  Note that as seen in Figure~\ref{fig:Case1Example6}, the crossover between the low $\tan\beta$ range and this intermediate to high $\tan\beta$ range occurs at quite modest values of $\tan\beta$ ($\sim 10$ and above); hence, the results shown here for $\tan\beta=20$ are quite similar to those at higher values of $\tan\beta$ (hence these higher values are not displayed separately).   In addition, the NLSP mass scales trivially with $\Lambda$, and thus as we've seen, the preferred Higgs mass region is associated with heavier neutralinos (and other superpartner masses) for low $\tan\beta$.

\begin{figure}[htbp]
   \centering
   \includegraphics[width = 0.65\textwidth]{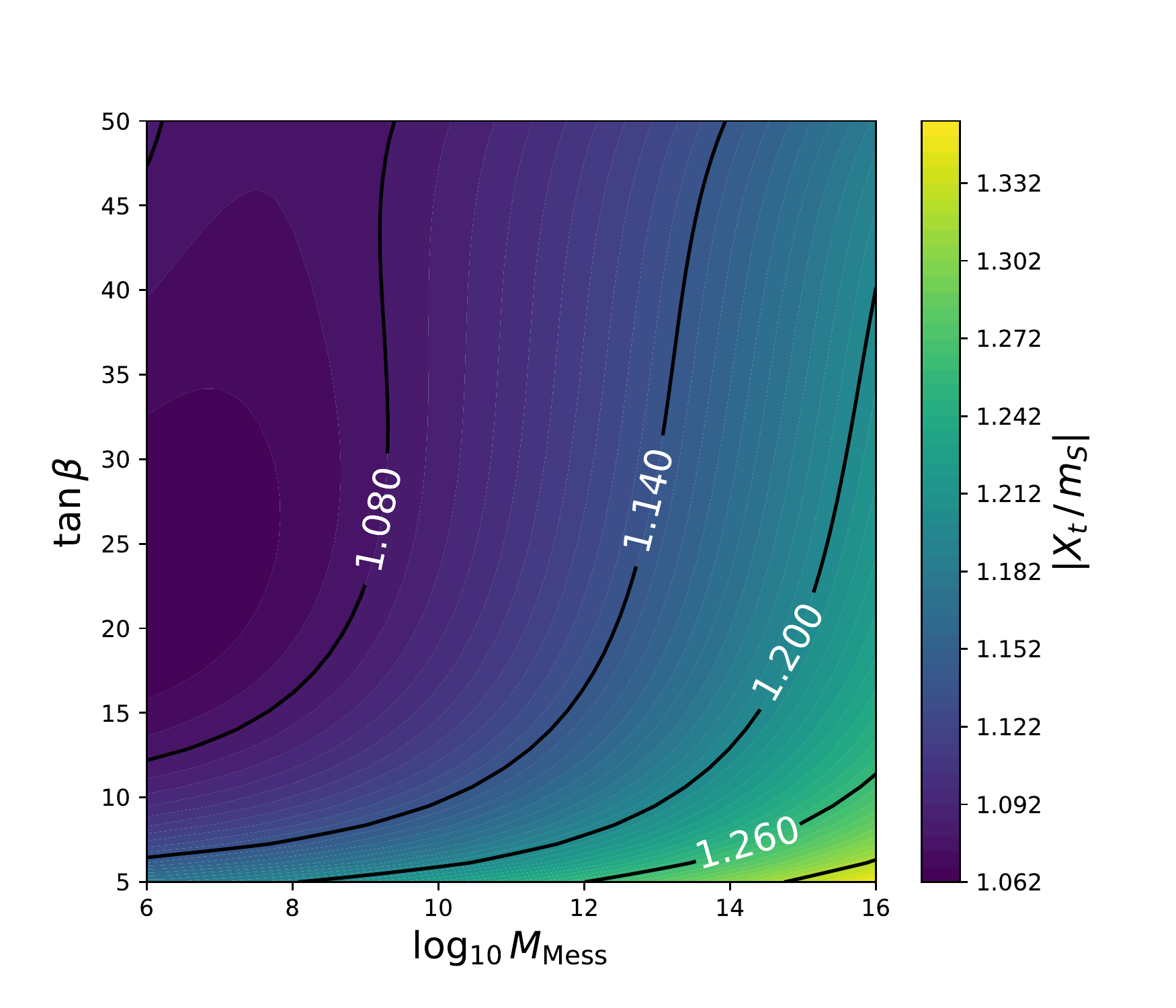} 
   \centering
      \caption{The values of $\vert X_t/m_S\vert$ in this scenario, where $X_t$ is the stop mixing parameter, and $m_S=\sqrt{m_{\tilde{t}_1}m_{\tilde{t}_2}}$, as displayed in the $\log M_{\rm mess}-\tan\beta$ plane.}
   \label{fig:stopmixing}
\end{figure}
It is illustrative to examine the stop mixing parameter $X_t=\tilde{A}_t-\mu\cot\beta$.  In Figure~\ref{fig:stopmixing}, we plot $\vert X_t/m_S\vert$ as a function of the messenger scale and $\tan\beta$, where the mass scale $m_S$ is given by the geometric mean of the stop masses ($m_S=\sqrt{m_{\tilde{t}_1}m_{\tilde{t}_2}}$). We see that we obtain sizable stop mixing throughout the parameter space in this scenario, with higher values in the case of high messenger scales, and lower values for low messenger scales, consistent with the mass spectra shown earlier.

We close this section by commenting that the superpartner mass range in this scenario is generally heavier than what is possible in flavored gauge mediation models based on $U(1)$ symmetries.  This can be seen for example for some of the benchmark points of \cite{Ierushalmi:2016axs}, for which all superpartner masses can be at the 2 TeV range or less.  In such models, the $U(1)$ charges can be chosen such that problematic couplings between the electroweak Higgs fields and the supersymmetry breaking field can be forbidden, whereas for non-Abelian Higgs-messenger symmetries, these couplings cannot be avoided.  Therefore, in the non-Abelian case, we need to augment the field degrees of freedom to ameliorate the effects of these couplings and arrive at a phenomenologically acceptable model.  As stated previously, this is the reason why we have $N_5\geq 2$ in the non-Abelian case.  For Abelian models, $N_5$ is a parameter that can be chosen, and thus one can obtain light spectra with $N_5=1$.

\section{Conclusions}
In this paper, we have explored a specific flavored gauge mediation model of the MSSM soft supersymmetry breaking parameters that can result from postulating that the electroweak Higgs fields and the $SU(2)$ messenger doublets are related by a  discrete non-Abelian gauge symmetry.  This Higgs-messenger symmetry is taken to be $\mathcal{S}_3$, as first studied in this context for two families in \cite{Perez:2012mj}, and later extended to three families in \cite{Everett:2016meb}. The model predicts two pairs of messenger fields, which transform as $\mathbf{5}$, $\bar{\mathbf{5}}$ representations of $SU(5)$; as discussed in \cite{Everett:2016meb}, this arises from the need to have an enlarged Higgs-messenger field content that includes $\mathcal{S}_3$ doublet and singlet representation to mitigate an otherwise severe $\mu/B_\mu$ problem.  The extended Higgs-messenger sector allows for a rich variety of possible renormalizeble superpotential couplings of the Higgs-messenger fields to the SM matter fields, depending on the assumed $\mathcal{S}_3$ charges of the quark and lepton superfields.  In a specific limit in which these couplings are dominated by the interactions among the $\mathcal{S}_3$ singlet representations, the resulting SM and messenger Yukawas both involve only third generation fields.  

As a result, a minimal flavored gauge mediation model is obtained in which the sfermion masses are flavor diagonal, and there is sizable stop mixing due to the one-loop third generation $A$ terms that arise from the messenger-matter interactions.  The model has three continuous parameters: the messenger scale $M_{\rm mess}$, the scale $\Lambda$, which sets the scale of the soft supersymmetry breaking terms (together with loop factors), and $\tan\beta$, and one discrete parameter (the sign of the $\mu$ parameter), which yields a highly predictive scenario. We showed in this paper that in much of the parameter space, the superpartner masses are at most $5-6$ TeV, with the gluino typically in the $4-5$ TeV range.  The exceptions to this general pattern occur at small values of $\tan\beta$, for which the need for large radiative corrections to bolster the light Higgs mass requires much heavier squark masses.  This highly predictive model of the MSSM soft parameters is thus one to keep in mind as the LHC continues to probe the paradigm of TeV-scale supersymmetry.

While the $\mathcal{S}_3$ singlet-dominated regime can arise trivially by requiring that the MSSM matter fields are inert to the Higgs-messenger symmetry, we have shown in this paper that it can also result in a specific limit in the case that the quark and lepton superfields have nontrivial $\mathcal{S}_3$ charges.  Hence, it is possible in this limit to sidestep the previously established correlation between the SM and messenger Yukawa couplings in this class of models \cite{Perez:2012mj,Everett:2016meb}, which had disallowed sizable third generation couplings for both the SM and messenger couplings, and consequently required heavier superpartner masses to obtain the experimentally determined value of the light Higgs mass. 

It is important to note that in the intriguing case that the quark and lepton superfields transform nontrivially with respect to the Higgs-messenger symmetry, reaching this limit requires additional symmetries that are also directly connected to the origin of the SM fermion masses and mixing parameters.  Indeed, given that in the scenario studied here, the first and second matter fermion generations are massless, extending this simple model to a fully realistic theory requires a detailed examination of subleading corrections as this limit is relaxed.  These corrections will yield flavor-changing interactions that are a hallmark of this class of flavored gauge mediation models. Though it is known that these effects can often be more strongly suppressed in flavored gauge mediation models than what naive estimates might suggest~\cite{Calibbi:2014yha,Ierushalmi:2016axs}, the question remains open as to whether a viable, fully-fledged three-family model can be constructed in which the Higgs-messenger symmetry is also a nontrivial part of the full family symmetry.  Further explorations along these lines are in progress \cite{fgmpaper3}.

\begin{acknowledgments}

We thank B.~Allanach for helpful discussions.  L.L.E. also thanks M.~McNanna for his work on this project at its early stages, and D.~Chung for constructive suggestions. T.S.G is thankful for S.~Guttman and her infinite patience.  This work is supported by the U. S. Department of Energy under the contract DE-SC0017647.
\end{acknowledgments}

\end{document}